\documentclass[12pt,amsmath,amssymb,aps,nofootinbib]{revtex4-1}
\usepackage{lipsum}
\usepackage{graphicx,epsfig,dcolumn,bm,epic,eepic,float}% Include figure files
\usepackage[T1]{fontenc}
\usepackage[utf8]{inputenc}
\usepackage{mathtools} 
\usepackage{pifont}
\usepackage{amsmath}
\usepackage{latexsym}
\usepackage{color}
\usepackage{cancel}
\usepackage{scrextend}
\usepackage{amsmath}
\usepackage{amsthm}
\usepackage{eucal}
\usepackage{amssymb}
\usepackage{mathrsfs}
\usepackage[bottom]{footmisc}
\usepackage{makeidx,shortvrb,latexsym}
\usepackage{epstopdf}
\usepackage{mathtools}
\usepackage{ragged2e}
\usepackage[T1]{fontenc}
\usepackage{lipsum}
\usepackage{array}
\usepackage{hyperref}
\usepackage[flushleft]{threeparttable}
\usepackage{longtable}
\usepackage{tasks}
\usepackage{blindtext}
\usepackage{enumitem}
\usepackage{multirow}
\usepackage{mathtools}
\newcommand {\be}{\begin{equation}}
\newcommand {\ee}{\end{equation}}
\newcommand {\ba}{\begin{eqnarray}}
\newcommand {\ea}{\end{eqnarray}}
\newcommand {\bea}{\begin{eqnarray}}
\newcommand {\eea}{\end{eqnarray}}

\pdfoutput=1

\numberwithin{equation}{section}
\begin{document}
\title{ CP violation in the extension of SM with a complex singlet scalar and vector quarks}
\author{\large Neda Darvishi$^{a,b,}$} \email{neda.darvishi@manchester.ac.uk}
\author{Maria Krawczyk$^{b,}$}\email{Deceased}

\affiliation{~~~~~~~~~~~~~~~~~~~~~~~~~~~~~~~~~~~~~~~~~~~~~~~~~~~~~~~~~~~~${}$\vspace{-3mm}\\
$^a$Consortium for Fundamental Physics, School of Physics and
  Astronomy,\\University of Manchester, Manchester M13 9PL, United
  Kingdom
  \\
 $^b$ Faculty of Physics,
University of Warsaw, 
Pasteura 5, 02-093 Warsaw, Poland}

\begin{abstract}
${}$

\centerline{\bf ABSTRACT} \medskip

\noindent
 We consider the simplest extension of the SM with a complex singlet and a pair of heavy doublet vector quarks, the so-called cSMCS model.
In this model, the CP violation can emerge spontaneously as a consequence of the time-dependent phase of the complex singlet vacuum expectation value and the mass mixing of the SM and heavy vector quarks. In our model, the CP-violating time-dependent phase depends on the Higgs field within the bubble-wall via the Higgs-singlet coupling that can directly explain the observed baryon-to-entropy ratio $\sim 9\times 10^{-11}$. Additionally, as a result of the mixing between vector quarks and SM quarks, the tree level Flavour Changing Neutral Currents and the charged-current decay channels $t' \to Wb, Zt, h_{i} t$ and $b' \to Wt, Zb, h_{i} b$ would arise that are well constrained by experimental data.  We investigate the implications of these constraints on the total cross-section for the production of heavy vector quark pairs via $pp \to t'\bar{t'}$ and $pp \to b'\bar{b'}$ channels and obtain the bounds on the heavy quark masses for this model.  Accordingly, we show the contribution of these heavy quarks in the Higgs bosons signal strength, $\mathcal{R}_{\gamma\gamma}$, as well as corrections to the gauge boson propagators.


\end{abstract}

\maketitle

\section{Introduction}
\label{Introduction}

The discovery of the Higgs particle at the CERN Large Hadron Collider
(LHC) \cite{Aad:2012tfa,Chatrchyan:2012xdj}, that was previously
predicted by the Standard Model (SM) of Particle Physics
\cite{Glashow:1961tr,Weinberg:1967tq,Salam1968,Englert:1964et,Higgs:1964pj},
generated more interest in beyond the SM Higgs Physics. This
is corroborated by the fact that the SM fails to address several key questions, such as the origin of the observed matter-antimatter asymmetry and the dark matter relic abundance in the Universe.
Particularly, understanding the origin of CP violation is one of the essential requirements to discover the mystery of the baryon asymmetry of the Universe~\cite{sak}. 
In the SM, CP symmetry is explicitly broken at the Lagrangian level through the complex Yukawa couplings. The main contribution of CP violation in this model can be parameterized by the Jarlskog invariant~\cite{Jarlskog}, which contains a single non-vanishing phase in the Cabibbo-Kobayashi-Maskawa (CKM) matrix~\cite{Cabibbo:1963yz,CKM}. However, the amount of CP violation in this model is inadequate to explain the observed baryon asymmetry of the Universe~\cite{Gavela:1994ds,Gavela:1994dt,ber}.

There are plenty of New Physics models with
additional Higgs scalars that have been introduced to provide additional sources of CP violation~\cite{comelli,Accomando:2006ga,Haber:2012np,Ginzburg:2004vp,Gunion:2005ja,Davidson:2013psa}. In this context, CP symmetry in the scalar sector might also have the possibility to be conserved at the Lagrangian level and spontaneously violated by the vacuum expectation value (VEV)~\cite{Lee1}. This kind of CP violation would play a role in multi-Higgs-doublet models and models with an additional singlet scalar~\cite{Lee1,Enqvist,Branco:1985pf,Branco:1985ch,Lavoura:1994fv,Botella:1994cs,MCDonald,Lavoura:1994yu,Branco:2003rt,Profumo:2007wc,Sokolowska:2008bt,AlexanderNunneley:2010nw,Lebedev:2010zg,Espinosa:2011eu,Gabrielli:2013hma,marco,Kozaczuk,Jiang:2015cwa,Barger:2008jx,Kozaczuk,Bonilla:2014xba,Krawczyk:2015xhl,Darvishi:2016tni,Darvishi:2017fwr,Darvishi:2019dbh,Birch-Sykes:2020btk}.

Here, we consider the simplest extension of the SM with a complex singlet in the presence of weak doublet heavy vector quarks, the so-called cSMCS model. In this model, there are three neutral Higgs particles, $h_{i}$, where the SM-like Higgs boson originates mostly from the SU(2)$_L$ doublet with a small correction arising from the singlet. 
In our framework, space- and time-dependent CP-violating phase can arise from the complex singlet VEV at high temperatures. This phase varies with the evolution of the Higgs field within the bubble-wall via the Higgs-singlet and self-singlet couplings. 
The contribution of this CP-violating phase in the mass mixing of the SM and heavy vector quarks results in the emergence of an additional source of CP violation that is spontaneous in nature. This source of CP violation directly explains the observed baryon-to-entropy ratio, $n_B/s\approx 9\times 10^{-11}$ ~\cite{WMAP,Ade:2013zuv}. Simultaneously, this model can provide a strong enough first-order electroweak (EW) phase transition as a necessary condition for a successful baryogenesis process~\cite{sak,Darvishi:2016tni,Darvishi:2017fwr}.

In this framework, due to the mixing between the SM quarks and the heavy vector quarks, flavour-changing~neutral~currents~(FCNCs)~\cite{Aguila} and charged currents are induced by $t' \to Wb, Zt,h_{i} t$ and $b' \to Wt, Zb, h_{i} b$ channels, which are well constrained by experimental data.  We investigate the implications of these constraints on the total cross-section for the production of heavy vector quark pairs via $pp \to t'\bar{t'}$ and $pp \to b'\bar{b'}$ channels to obtain the bounds on the masses of vector quark for this model. We then use these bounds to compute the additional contributions from heavy vector quarks in Higgs bosons production via $pp \to t'\bar{t'}/b'\bar{b'}\to h_i X$ channels and $h_i\to \gamma \gamma$ decays in the Higgs bosons signal strength, $\mathcal{R}_{\gamma\gamma}$. Additionally, we parametrize the corrections to the gauge boson propagators induced by new Higgs bosons and vector quark in our model. 
Remarkably, the proprieties of the SM-like Higgs particle in this model is in an excellent agreement with recent LHC measurements, including its production cross-section that can be fitted to the experimental data from ATLAS and CMS at $2\sigma$ confidence level (CL)~\cite{Darvishi:2016fwo}. 

Furthermore, this model is a part of a larger framework introduced in \cite{Bonilla:2014xba,Krawczyk:2015xhl}, where the extension of the SM by a complex singlet and the inert doublet has been studied, with the focus on the properties of dark matter. 
 
The layout of this paper is as follows. In Section \ref{sec:1}, we describe the basic features of cSMCS model. In Section \ref{sec:2}, we discuss the introduction of CP violation via the CP-violating phase originated from VEV of the complex singlet, including the contribution of this phase in the mixing of SM quarks and heavy vector quarks. The physical states in the Higgs sector will be discussed in Section \ref{sec:3}. This section also contains the results of our scanning over the parameters of the model in the region of the non-vanishing phase and in alignment with the SM-like Higgs boson mass limit. In Section \ref{sec:vq}, we analyse the properties of vector quarks by exploring signals arising from  $pp \to t'\bar{t'}$ and $pp \to b'\bar{b'}$ production and $t' \to Wb,Zt,h_{i} t$ and $b' \to Wt,Zb,h_{i} b$ decays channels. Furthermore, we find the mass spectrum of these heavy particles within our framework. In Section \ref{sec:4}, we present our numerical estimates for the Higgs bosons signal strength, $\mathcal{R}_{\gamma\gamma}$, as well as corrections to the gauge boson propagators. Section \ref{sec:con} contains our remarks and conclusions. 
Finally, technical details are delegated to Appendices \ref{appA}, \ref{hhh} and \ref{A-obli}.

\section{The cSMCS model: The SM plus a complex singlet and heavy doublet vector quarks}
\label{sec:1}

The Higgs sector of the cSMCS model may be described by a scalar doublet and a complex singlet scalar, as
\begin{equation}
{\Phi}\ =\ \left(
 \begin{matrix}
\phi^+\\
 \phi^0
 \end{matrix}\right)\;,\qquad 
\chi\;.
\end{equation} 
Observe that 
$\Phi$ transforms covariantly under an SU(2$)_L$
gauge transformation as
\begin{equation}
\Phi\to \text{U}_L\Phi, \quad \text{U}_L\in \text{SU(2)}_L.
\label{su(2)}
\end{equation}

The SU(2)$_L\otimes$U(1)$_Y$-invariant Lagrangian is given by
\begin{equation}
{ \cal L}={ \cal L}^{}_{ gf } + {\cal L}_Y(\psi_f,\Phi) + {\cal L}_Y(\psi_f,\chi)+{ \cal L}_{scalar} ,  \,, \quad { \cal L}_{scalar}=T-V,
\label{lagrbas}
\end{equation}
where ${\cal L}^{}_{gf}$ stands for the gauge bosons interactions with fermions. The ${\cal L}_Y(\psi_f,\Phi)$ and ${\cal L}_Y(\psi_f,\chi)$ show the Yukawa sector of the model that contains interactions of scalar doublet and singlet scalar with vector quarks and SM quarks. 
Here, we consider a pair of doublet vector quarks $Q^4_{L}\equiv V_{L}=(t',\,  b')_L$ and $V_R=(t',\,  b')_R$ with the properties similar to $Q_L$ in the SM, with ${\rm U(1)}_Y$-hypercharge $Y=1/6$. 
Finally, the last term, ${ \cal L}_{scalar}$, shows the scalar sector of the model.

The gauge-kinetic term in ${ \cal L}_{scalar}$ takes the form,
	\begin{eqnarray}
	T = \left( D_{\mu} \Phi \right)^{\dagger} \left( D^{\mu} \Phi \right) + \partial \chi \partial \chi^*. \label{kinet}
	\end{eqnarray}
The covariant derivative $D_{\mu} $ for an SU(2)$_L$ doublet may be defined as
	\begin{equation}
 D_{\mu}\ =\ \sigma^0 \partial^0_{\mu} +i g_w W_{\mu}^i {\sigma^i \over 2}\:
+\: i{g_Y\over 2} B_{\mu} \sigma^0,
\end{equation}
where $\sigma^{1,2,3}$ are the Pauli matrices.

The SU(2)$_L\otimes$U(1)$_Y$-invariant Higgs potential is given by
\begin{equation}
V=V_{D}+V_S+V_{DS}, \label{potgen}
\end{equation}
where $V_{D}$ stands for pure doublet interaction similar to the SM potential as
\begin{equation}
	V_{D} = - \frac{1}{2}{m_{11}^2} \Phi^\dagger\Phi 
	+ \frac{1}{2}\lambda \left(\Phi^\dagger\Phi\right)^2.
\label{potSM}
\end{equation}
The $V_{S}$ contains the self-interaction of the complex singlet, which takes on the form
\begin{eqnarray}
V_{S} =&& -\frac{1}{2}m_{s1}^2 \chi^* \chi -\frac{1}{2} m_{s2}^2 (\chi^{*2} + \chi^2)
\nonumber\\&&
+ \lambda_{s1} (\chi^*\chi)^2 + \lambda_{s2} (\chi^*\chi)(\chi^{*2} + \chi^2) + \lambda_{s3} (\chi^4 + \chi^{*4})
\nonumber\\&&
+ \kappa_1 (\chi + \chi^*) + \kappa_2 (\chi^3 + \chi^{*3}) + \kappa_3(\chi^*\chi)(\chi+ \chi^*). 
\label{potS}
\end{eqnarray}
Finally, the doublet-singlet interaction $V_{DS}$ is,
\begin{eqnarray}
V_{DS} = &&\Lambda_1(\Phi^\dagger\Phi)(\chi^* \chi) + \Lambda_2 (\Phi^\dagger\Phi)(\chi^{*2}+ \chi^2)
\nonumber\\&&
+ \kappa_4 (\Phi^\dagger\Phi) ( \chi +\chi^*). 
\end{eqnarray}
Hence, the full potential introduces three mass terms ($m^2_{11},\, m^2_{s1}, \, m^2_{s2}$), six dimensionless quartic ($\lambda,\lambda_{s1},\lambda_{s2},\lambda_{s3}, \Lambda_1, \Lambda_2$) and four dimensionful parameters ($\kappa_{1},\kappa_{2},\kappa_{3},\kappa_{4}$). Thereby, the potential $V$ contains 13 real parameters and is invariant under $\chi \to \chi^*$.
	
Here, we consider both scalar fields $\Phi$ and $\chi$ to receive non-zero VEVs. Following the standard linear expansion of  the two scalar fields about their VEVs, we may re-express them as
\begin{equation}
\Phi = \left( \begin{array}{c} G^{\pm} \\ \frac{1}{\sqrt{2}} \left( v + \phi_1 + i G^0 \right)\\ \end{array} \right), \quad
\chi = \frac{1}{\sqrt{2}} (w\, e^{i \xi} + \phi_2 + i \phi_3), 
\label{dec_singlet}
\end{equation}
with complex VEV for complex singlet with constant phase at $T = 0$. The phase of the VEV of complex singlet field may be then responsible for the spontaneous CP violation. 

After spontaneous symmetry breaking (SSB), the standard EW gauge fields,
the $W^{\pm}$ and $Z$ bosons, acquire their masses from the three
would-be Goldstone bosons $(G^{\pm},G^0)$~\cite{Goldstone}. Consequently, 
the model has only three physical scalar states; two
CP-even scalars ($h_1$,$h_2$) and one CP-odd scalar $(h_3)$. 

Having defined the potential, let us try to simplify the model with the help of
symmetry transformations that would leave the model
invariant. These symmetries impose constraints over the theoretical
parameters of the models and thus enhance their predictability. In this model, several symmetries can be realized including a global U(1)-symmetry and three discrete symmetries, i.e. $Z_2,\, Z_3$, and $Z_4$. Note that the potential originally is invariant under the CP transformations; $\Phi \to \Phi^*$ and $\chi \to \chi^*$.
Under the global U(1)-symmetry and the Abelian discrete symmetry group, $Z_n = \{1,\omega_n,\dots,(\omega_n)^{n-1}\}$, 
the scalar fields transform as
\begin{align}
  {\rm	U(1)}:& \qquad\Phi \to \Phi\,, \qquad \chi\to e^{i\alpha} \chi,
\end{align}
and
\begin{align}
     Z_n:& \qquad \Phi\to\Phi\,, \qquad \chi\to \omega_n \chi\,,
\end{align}
with $(\omega_n)^n=1$. The non-zero parameters corresponding to cSMCS potentials which are invariant under these symmetries are given in Table \ref{tab:sym}.

\begin{table}[h]
\begin{tabular}[t]{ | c | l | }
\hline
  \small{Symmetry}  &  \small{\,Non-zero Parameters}  \\
\hline \hline

$Z_2$ &  \begin{tabular}{@{}l@{}} $\,m_{11}^2,\, m_{s1}^2,\,m_{s2}^2,\, \lambda,\, \lambda_{s1},\,\lambda_{s2},\,\lambda_{s3},\, \Lambda_{1}, \, \Lambda_{2}$ \end{tabular}  \\ \hline

 $Z_3$ & \begin{tabular}{@{}l@{}}  $\,m_{11}^2,\, m_{s1}^2,\, \lambda,\, \lambda_{s1},\,\lambda_{s3},\, \Lambda_{1},\, \kappa_2,\, \kappa_3$ \end{tabular}  \\ \hline

 $Z_4$ & \begin{tabular}{@{}l@{}} $\,m_{11}^2,\, m_{s1}^2,\, \lambda,\, \lambda_{s1},\,\lambda_{s3},\, \Lambda_{1}$  \end{tabular}  \\ \hline

 U(1) & \begin{tabular}{@{}l@{}}  $\,m_{11}^2,\, m_{s1}^2, \,\lambda_{1}, \,\lambda_{s1},\, \Lambda_{1}$
 \end{tabular}  \\ \hline
\end{tabular}
\caption{\it Non-zero parameters for the symmetric cSMCS potentials}
\label{tab:sym}
\end{table}

Meanwhile, to have spontaneous CP violation at high temperature, the CP-violating phase, $\xi$, needs to proportional to the Higgs field within the bubble-wall. This demand can be met either by $Z_2$ symmetry or by broken-U(1) symmetry. Besides, in the case of U(1) symmetry, a non-zero VEV of singlet would break U(1) symmetry and massless Nambu-Goldstone scalar particles appear in the model. Assuming some U(1) soft-breaking terms in the potential, that still leads to a fewer number of parameters, would help to avoid these massless particles. 

In this study, we shall focus on potential with a soft-breaking of U(1) symmetry, where the singlet cubic terms $\kappa_{2},\;\kappa_{3}$ and the singlet quadratic term $m^2_{s2}$ are preserved. Note that the linear term $\kappa_1$ can be rotated away by a translation of the complex singlet field. Also, the $\kappa_4$ term is negligible since it may be generated at one loop with the strength given by $\frac{1}{16 \pi^2} \kappa_3 \Lambda_1$, where coupling $\Lambda_1$ is small to ensure perturbativity of the model. Therefore, the model contains the U(1)-symmetric terms ($m_{11}^2, m_{s}^2, \lambda, \lambda_{s1}, \Lambda_{1}$) and the U(1)-soft-breaking terms ($m_{s2}^2,\kappa_{2},\kappa_{3}$). Hence, the potential takes on the following form
\begin{align} 
V =& -\frac{1}{2}{m_{11}^2} \Phi^\dagger\Phi+ \frac{1}{2}\lambda\left(\Phi^\dagger\Phi\right)^2 
\nonumber\\&
- \frac{1}{2}{m^2_{s1}} \chi^* \chi + \lambda_{s} (\chi^*\chi)^2 + \Lambda(\Phi^\dagger\Phi)(\chi^* \chi)
\nonumber\\&
 - \frac{1}{2}{m^2_{s2}} (\chi^{*2} + \chi^2)+ \kappa_2 (\chi^3 + \chi^{*3}) + \kappa_3 (\chi^*\chi)(\chi + \chi^*), 
\label{pot-cons}
\end{align}
with $ \lambda_s= \lambda_{s1}$ and $\Lambda=\Lambda_1$ and all real parameters. 

In the next section, we discuss the introduction of CP violation via the mixing of SM quarks and heavy vector quarks due to complex VEV of the complex gauge singlet scalar.

\section{Spontaneous CP VIOLATION in cSMCS}\label{sec:2}
\begin{figure*}[t]
\includegraphics[width=.6\textwidth]{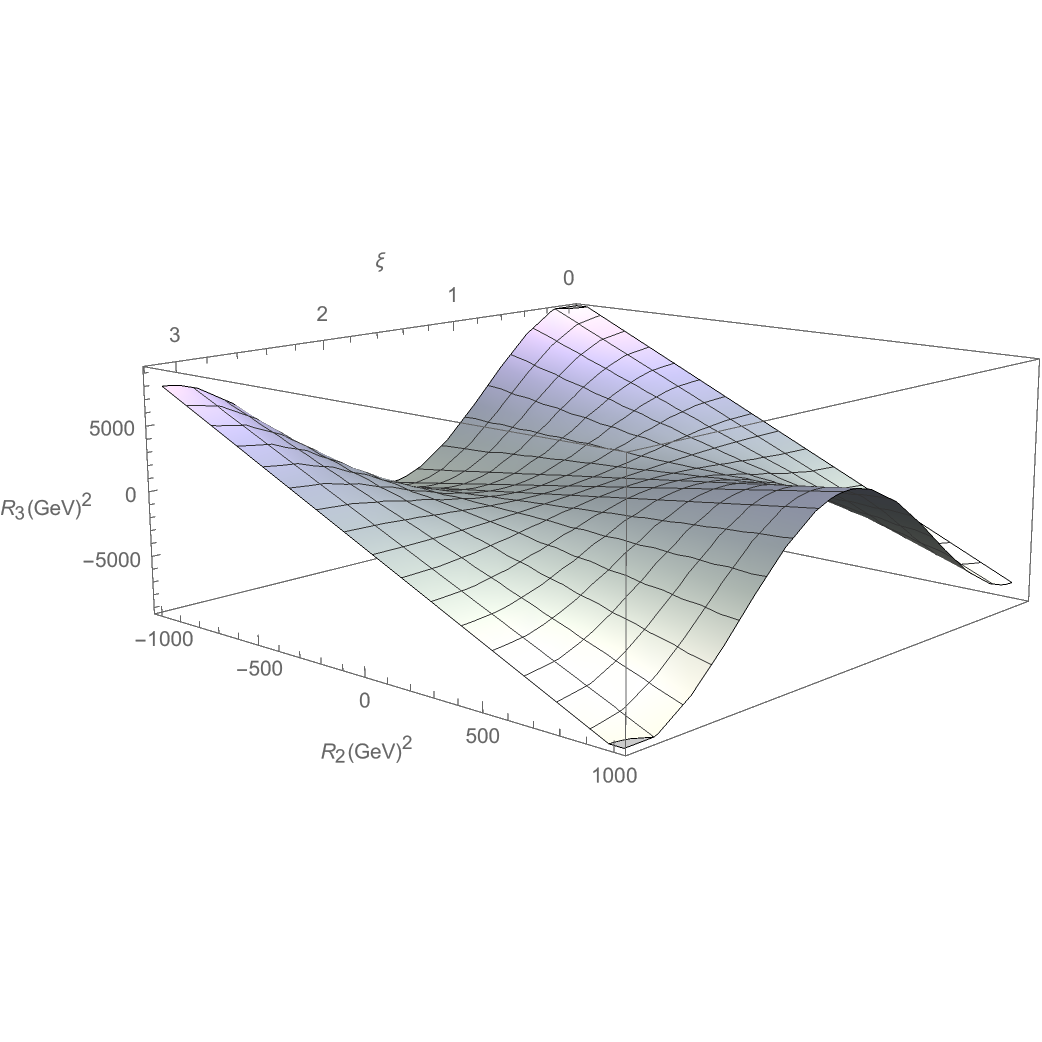}\\
\includegraphics[width=.8\textwidth]{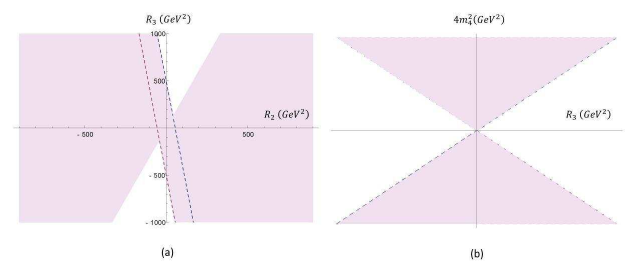}
\caption[CP violation]{\it The correlation between the parameters $R_2$, $R_3$ and $m^2_{s2}$,
for $-1 < \cos 2\xi < 1$ with $R_2=\sqrt{2}w\kappa_2 $ and $R_3=\sqrt{2}w\kappa_3$. The dashed lines correspond to the limit when phase vanishes. }
\label{fig1}
\end{figure*}

Having introduced non-zero VEVs for $\langle \Phi\rangle= v/\sqrt{2}$
and $\langle \chi\rangle= w e^{i \xi}/\sqrt{2}$, we may obtain minimization conditions of potential ~\eqref{pot-cons} with respect to $v$, $w$ and $\xi$, as
\begin{align}
&v \left(-m_{11}^2+\lambda \, v^2+ \Lambda\, w^2\right) =0, 
\\
&w \left(- m_{s1}^2-2\, m_{s2}^2 \cos 2 \xi +\Lambda\, v^2+2\,
   \lambda_s w^2+3 \sqrt{2} \, w \, \left(\kappa_2 \cos 3 \xi +
   \kappa_3 \cos \xi \right)\right)=0,
\label{min2} \\
&w^2 \left(\sqrt{2}\, m_{s2}^2  \sin 2 \xi -w \left(3\, \kappa_2 \sin 3 \xi+
 \kappa_3 \sin \xi \right) \right)=0,
\label{min3} 
\end{align}	
and hence, to have a non-vanishing phase the follwoing relation must be satisfied~\cite{Krawczyk:2015xhl}
	\begin{equation}
	- 4 m^2_{s2} \cos\xi + \sqrt{2}w \big( 3\kappa_2 (1+2\cos2\xi)+\kappa_3\big)=0.
	\label{CP}
	\end{equation}
Observe that the phase $\xi$ depends on the cubic parameters $\kappa_2$ and $\kappa_3$.
Evidently, considering the limit $\kappa_2\to 0$ in the Eqs.~\eqref{min2} and \eqref{min3} for a given $v$, we may obtain
\begin{align}
\cos\xi =  {\sqrt{2}  \, \kappa_3 \, w\over 4 m_{s2}^2},
\label{cos}
\end{align}
and
\begin{align}
w^2 =  {m_{s1}^2-\Lambda\, v^2-2\, m_{s2}^2 (3+2\cos 2\xi )\over  2 \lambda_s },
\end{align}
which are true as long as 
$|{\sqrt{2} \, \kappa_3 \, w / 4 m_{s2}^2}|<1$. 
In the same way, for the limit $\kappa_3\to 0$ we find a non-vanishing phase within the boundary
$|(m_{s_2}^2 \pm \sqrt{m_{s_2}^4 + 18 w^2 \kappa_2^2})/6\sqrt{2}w\kappa_2|<1$. 
Hence, only one of the cubic terms could be enough to have a non-vanishing phase. Nevertheless, to maintain generality, we will consider both parameters in our calculations. 
Figure~\ref{fig1} shows the correlation between cubic couplings $\kappa_2$, $\kappa_3$ in the $|\cos 2 \xi|< 1$ plane, by considering $4 m^2_{s2}=500\,$ GeV$^2$.  
%Although, instead we can also consider other couplings (such as $\lambda_{s3} (\chi^4 + \chi^{*4})$ that could contribute for non vanishing phase, but for now we only consider one of the cubic terms which are the important ones for EW baryogenesis. 

Now, let us discuss how the CP-violating phase would contribute to CP violation.
Obviously, both $w$ an $\xi$ are dependent on the Higgs field within the bubble-wall via the coupling $\Lambda (\Phi^\dagger\Phi)(\chi^* \chi) $.
During the EW phase transition, $v(x)$ is space- and time-dependent and hence $w(x)$ an $\xi(x)$ are space- and time-dependent. On the other hand, the Yukawa interaction of the model is given by
\begin{equation}
 {\cal L}_Y= y_{ij}^u \overline{Q}^i_L \widetilde{\Phi} \, {u^j_R} + y_{ij}^d \overline{Q}^i_L \Phi \, {d^j_R}+ \lambda_V \chi \overline{Q}^j_L V_R + M \overline{V}_L V_R+ h.c.,
\label{LY}
\end{equation}
with $i=1,\cdots, 4$; $j=1,2,3$ and $Q^4_{L}\equiv V_{L}$. The Yukawa couplings through Higgs doublet $\Phi$ denoted by $y_{ij}$ and through singlet denoted by $\lambda_V$ and $M$ is the mass of vector quarks. 

 By assuming the most dominant contribution of SM fermions (i.e. terms involving the heaviest quark generation in the SM) in the presence of the space- and time-dependent Higgs field in the bubble-wall, we may obtain the following transformation for $Q$ and $V$
\begin{align}
	\begin{pmatrix}
    Q^{'}_{L} \\
    V^{'}_{L} \\
	\end{pmatrix}
	&=
	\begin{pmatrix}
    A & B \\
    - B^{*} & A^{*} \\
	\end{pmatrix}
	\begin{pmatrix}
    Q_{L} \\
    V_{L} \\
	\end{pmatrix},
\end{align}
where
\begin{align}
&A= \left[ 1 + \left( {\lambda_V w \over M} \right)^2 \right]^{-1/2}, \nonumber \\
&B = \left( {\lambda_V w \over M} \right) \left[ 1 + \left( {\lambda_V w \over M} \right)^2 \right]^{-1/2} \;
        e^{-i\xi}.
\label{D}
\end{align}
Accordingly, by diagonalising the quark mass matrix, a couple of space- and time-dependent terms arise in the Lagrangian, i.e. 
\begin{eqnarray}
 \overline{Q}_L i \gamma^\mu \partial_\mu Q_L+&& \overline{V}_L i \gamma^\mu \partial_\mu V_L \to \overline{Q'}_L i \gamma^\mu \partial_\mu Q'_L+ \overline{V'}_L i \gamma^\mu \partial_\mu V'_L+ \Delta\mathcal{L}_k + const,
 \label{potsS}
 \end{eqnarray}
with
\begin{equation}
\Delta\mathcal{L}_k=-{\lambda_V^2 w(x)^2 \over M^2}\dot{\xi}(x)(\overline{Q'}_L \gamma^0 Q'_L-\overline{V'}_L \gamma^0 V'_L),
\end{equation}
where at $T = 0$, these terms disappear due to a constant phase in the mass term. 

Having obtained the above relation, the baryon asymmetry generated during the EW phase transition (in the absence of diffusion effects) then follows from the standard spontaneous baryogenesis analysis~\cite{5,6}.  This mechanism works when there is an interaction of the form
\bea
{\mathcal{L}}=\partial_\mu \xi J_B^\mu,
\eea
where $J_B^\mu$ is the baryon current. Therefore, the phase of the singlet VEV should be space- and time-dependent to generate baryon asymmetry.
The observations from WMAP gives the ratio of the baryon number to the entropy as 
${n_B/s}= (8.7\pm0.3) \times 10^{-11 }$ ~\cite{WMAP,Ade:2013zuv}.
Henceforth, in order to have successful baryogenesis $ {\lambda_V^2 w^2 \delta\xi / M^2 } =  (1.14\pm0.3) \times 10^{-3}$, while ensuring EW phase transition is strongly enough first order ~\cite{Darvishi:2016tni,Darvishi:2017fwr}.

Note that in addition to this source of CP violation, there is a $T = 0$ CP violation due to a constant phase in the mass term, that act as a CP-violating phase in the Kobayashi-Maskawa matrix. This will be shown in Section \ref{sec:vq}.

In the next section, we will discuss the allowed regions of parameters in the model for the non-vanishing phase and considering the limit for SM-like Higgs boson mass.

\section{Physical states in the Higgs sector}
\label{sec:3}

In this model there are three neutral Higgs bosons with mixed CP properties. Let us first introduce the Mass squared matrix, $M_{mix}$, in the basis of $\phi_1,\;\phi_2,$ and $\phi_3 $ as
\begin{equation}
M_{mix} = \left(
\begin{array}{ccc}
M_{11} & M_{12} & M_{13}\\
M_{21} & M_{22} & M_{23}\\
M_{31} & M_{32} & M_{33}
\end{array} \right),
\label{mixed}
\end{equation}
where the $M_{ij} (i,j=1,2,3)$ are,
\begin{equation}
 \begin{array}{l}
M_{11}=\frac{1}{2} \left( -m_{11}^2+3 v^2 \lambda+w^2 \Lambda\right),
\\[2mm]
M_{12}=v w_1\Lambda,
\\[2mm]
M_{13}=v w_2 \Lambda,
\\[2mm]
M_{22}=-m^2_{s2} - \frac{1}{2}{m^2_{s1}}+\frac{1}{2} v^2 \Lambda+(w_2^2 +3 w_1^2) \lambda_s +3\sqrt{2}w_1(\kappa_2+\kappa_3),
\\[2mm]
M_{23}=w_2\left(2 w_1 \lambda_s +\sqrt{2}(-3\kappa_2+\kappa_3)\right),
\\[2mm]
M_{33}=m^2_{s2} - \frac{1}{2}{m^2_{s1}}+\frac{1 }{2} v^2 \Lambda+(w_1^2 +3 w_2^2) \lambda_s+\sqrt{2}w_1(-3\kappa_2+\kappa_3),
 \end{array}
\label{masses}
\end{equation}
with $w_1=w\cos\xi$ and  $w_2=w\sin\xi$.

To obtain the masses of the three scalars in their physical basis $h_1,h_2$ and $h_3$, we need
to diagonalise the three-by-three mass matrix~$M^2_{mix}$
\begin{eqnarray}
 r M_{mix}^2 r^T = diag(M_{h_1}^2,M_{h_2}^2,M_{h_3}^2), \label{neutr_diag}
\end{eqnarray}
where the rotation matrix $r$ depends on three mixing angles ($\alpha_1, \alpha_2, \alpha_3$).
%\begin{eqnarray}
% & \left( \begin{array}{c} h_1\\ h_2\\ h_3\\ \end{array} \right) =r \left( \begin{array}{c} \phi_1\\ \phi_2\\ \phi_3\\ %\end{array} \right),
%\end{eqnarray}
The full rotation matrix $r$ takes on the form
\begin{eqnarray}
r =&&  \left(
\begin{array}{ccc}
c_1 c_2 & c_3 s_1 - c_1 s_2 s_3 & c_1 c_3 s_2 + s_1 s_3\\
-c_2 s_1 & c_1 c_3 + s_1 s_2 s_3 & -c_3 s_1 s_2 + c_1 s_3\\
-s_2 & -c_2 s_3 & c_2 c_3
\end{array} \right),
\label{rotfull}
\end{eqnarray}
where $s_i \equiv \sin \alpha_i$ and $c_i \equiv \cos \alpha_i$.
Hence, we may obtain
\begin{equation}
M_{h_1}^2 \simeq v^2 \lambda,
\end{equation}
\begin{equation}
M_{h_{2,3}}^2 =\frac{1}{2}(M_{22}+M_{33}\mp\sqrt{(M_{22}+M_{33})^2+4M_{23}^2}),
\end{equation}
where the $M_{h_1}$ corresponds to the SM-like Higgs boson that is identified with $\sim125$ GeV resonance, observed at the LHC \cite{Chatrchyan:2012xdj,Aad:2012tfa}. We take the masses of the two other Higgs bosons, $M_{h_2}$ and $M_{h_3}$, with the following heirarchy~\cite{Bonilla:2014xba},
\begin{equation}
M_{h_3} 
%>\lesssim
\gtrsim M_{h_2} > 150 \, {\rm GeV}.
\end{equation}

Now, to perform our scanning the following bounds are adopted 
\begin{align}
\quad &v \in [246,\,247] \; {\rm GeV}, &&M_{h_1} \in [124.5, 126]\; {\rm GeV},
 \nonumber \\[0.05in] 
\quad &\lambda_{1} \in[0.2, \,0.3], &&\Lambda \in[-1,\,  1],
\nonumber \\[0.05in]   
\quad &\rho_{2,3} \in[-1,\,  1],  && \xi \in [0,\, \pi],  
 \nonumber \\[0.05in]                                           
 \quad & m_{s1}^2, \, m_{s2}^2,\, m_{11}^2  \in [ -90000, 90000] \,\, {\rm GeV}^2,&& \lambda_{s} \in[0, \,1],
\end{align}
where we use the dimensionless parameters $\rho_{2,3}=\kappa_{2,3}/ w$.

%\begin{equation}
%\begin{array}{l}
% \lambda> 0, \,\, \lambda_{s} > 0, \quad
% \Lambda > - \sqrt{2 \lambda \lambda_{s}}.\\[3mm]
% \end{array} \label{pos}
%\end{equation} 

\begin{figure*}[t]
\includegraphics[width=.99\textwidth]{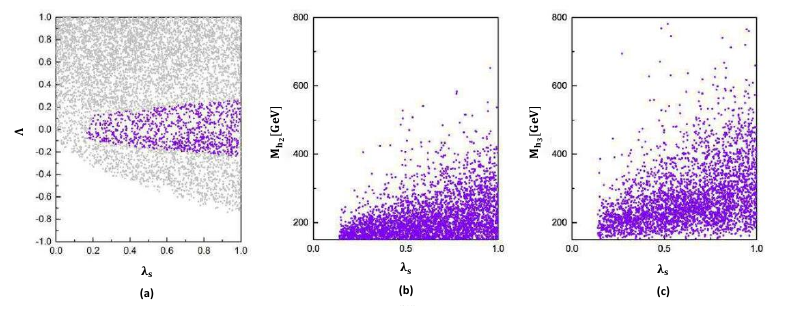}
\caption[]{\it The parameters relations of the model in 
(a) ($\lambda_s$,\, $\Lambda$), (b) ($\lambda_s$, $M_{h_2}$) and (c) $(\lambda_s$, $M_{h_3}$) planes.}
\label{fig3}
\end{figure*}

Figure \ref{fig3}(a) displays the correlation between parameters $\Lambda$ and $\lambda_s$. The grey region shows the constraint applied by positivity conditions and the violet region is the mass limit confinement. Notice that the mass limit confinement imposes a stronger constraint compared to the usual convexity conditions that are assumed to ensure the stability of the potential\footnote{However, the usual constraints derived from convexity conditions on quartic couplings for the stability of potential might be over-restrictive and unnecessary in some theoretical frameworks~\cite{Darvishi:2019ltl,Darvishi:2020teg}.}. From Figure~\ref{fig3}(a), it may be observed that the lower limit for singlet self-coupling is $\lambda_s\sim 0.2$, and the upper limit for doublet-singlet coupling is $|\Lambda|\sim 0.2$.  In Figure \ref{fig3}(b) and \ref{fig3}(c), we give our numerical estimates of the mass spectrum for heavy Higgs bosons ${h_2}$ and ${h_3}$. We may observe the
domains in which $\lambda_s$ becomes larger the masses $M_{h_2}$ and $M_{h_3}$ reach up to $\sim 650$\,GeV and $\sim 800$\,GeV.

\begin{figure*}[t]
\includegraphics[width=.99\textwidth]{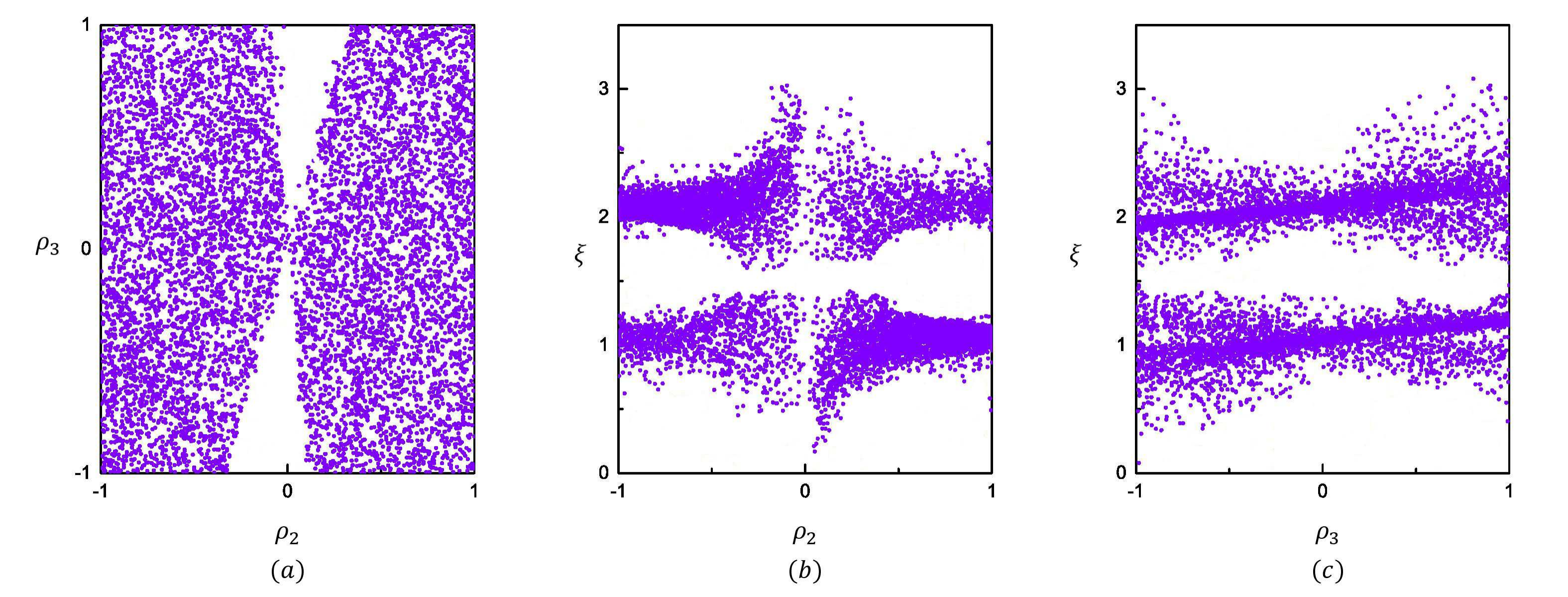}
\caption[]{\it The relations between cubic parameters in
(a) ($\rho_2$,$\rho_3$), (b) ($\rho_2$, $\xi$) and (c) ($\rho_3 $, $\xi$) planes.}
\label{fig4}
\end{figure*}

Figure~\ref{fig4} shows the correlation between cubic parameters $\rho_2$ and $\rho_3$ in the domain of the non-vanishing phase. By comparing Figure~\ref{fig4}(a) and Figure~\ref{fig1}, one may notice that the mass limit does not put an additional constraint over the cubic parameters. However, these parameters are directly confined by the CP-violating phase, $\xi$.

\begin{figure*}[t]
\includegraphics[width=.7\textwidth]{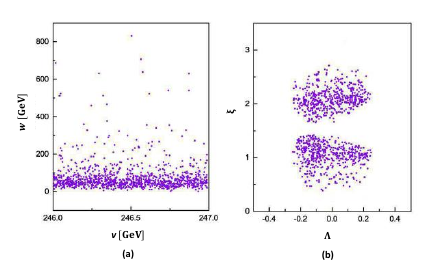}
\caption[]{\it The correlation between parameter of the model in
(a) ($v$, $w$) and
(b) ($\Lambda$, $\xi$) planes.} 
\label{fig5}
\end{figure*}
In Figure \ref{fig5}(a), we give our numerical estimates of the singlet's VEV, $w$, in the $(v,\,w)$ plane. Observe that $w$ may reach up to $\sim 800$~GeV and its lower limit is around $\sim 2$~GeV. Figure \ref{fig5}(b) shows the constraint imposed by the non-vanishing phase over $\Lambda$, which along with the aforementioned constraint imposed by mass limit, strongly restrict this parameter.

In the next section, we will analyse the properties of vector quarks by exploring signals arising from  $pp \to t'\bar{t'}$ and $pp \to b'\bar{b'}$ production channels and $t' \to Wb,Zt,h_{i} t$ and $b' \to Wt,Zb,h_{i} b$ decays~at~the~LHC. Comparing our result with the predictions from ATLAS and CMS may give us a sense of the mass spectrum of these heavy particles~\cite{Aaboud:2018pii,Sirunyan:2018qau}.

\section{Signatures of vector quarks at LHC}\label{sec:vq}
As we have shown in Section \ref{sec:2}, the mixing between the SM quarks and vector quarks is generated by the Yukawa interaction of the model. 
The mass matrices for this mixing may be given by
\begin{align}
M_{mix}^{u/d} = 
\begin{pmatrix}
m^{u/d}_{ij} & \mu_{ik} \\
\mu_{ki}     & M 
\end{pmatrix}, \quad i , j = 1,2,3, \quad k =4,
\end{align}
with $m^{u/d}_{ij} = y^{u/d}_{ij}v/\sqrt{2}$ and the mixing terms $\mu_{i4}=\lambda_V^i w e^{-i\xi}/\sqrt{2}$ and $\mu_{4i}=y_{i4}v/\sqrt{2}$.
%$$
%\mu^{iv}={1 \over \sqrt{2}} \bigg[ y^{ki} \sum_{l=1}^3r_{l1}+\lambda_V^{ik} 
%	\big( \sum_{l=1}^3r_{l2}-i\sum_{l=1}^3r_{l3} \big) \bigg].
%$$
The above mass matrices are diagonalized by making unitary transformations $V^{t,b}_L$ and $V^{t,b}_R$ so,
\begin{align}
&{V^{t}_L}^\dagger M_{mix}^{u} V^{t}_R= diag(m_{u},\, m_{c}, \, m_{t},\, M_{t'}), \label{top_diag}
\\
&{V^{b}_L}^\dagger M_{mix}^{d} V^{b}_R= diag(m_{b},\, m_{s}, \, m_{b},\, M_{b'}). \label{bottom_diag}
\end{align}
After rotating the weak eigenstates into the mass eigenstates, the Higgs bosons couplings to SM quarks and heavy vector quark pairs may be modified as
\begin{align}
 g^{h_i}_{q_j\bar{q_j}} &= r_{i1} \bigg[ (V^{q}_L)^{*}_{jj} (V^{q}_R)_{jj}  {\sqrt{2} m_{q_j} \over v } 
						+ (V^{q}_L)^{*}_{j4} (V^{q}_R)_{jj} { y_{j4}^{q} } 
						+ (V^{q}_L)^{*}_{jj} (V^{q}_R)_{4j} {(r_{i2} - i r_{i3}) \over r_{i1}} {\lambda_V^j }  \bigg],
 \nonumber \\
 g^{h_i}_{q_4\bar{q_4}} &= r_{i1} \sum_{q_j} \bigg[ (V^{q}_L)^{*}_{j4} (V^{q}_R)_{j4}  {\sqrt{2} m_{q_j} \over v }
						+ (V^{q}_L)^{*}_{44} (V^{q}_R)_{j4} { y_{j4}^{q} }
						+ (V^{q}_L)^{*}_{j4} (V^{q}_R)_{44} {(r_{i2}- i r_{i3}) \over r_{i1}} {\lambda_V^j }  \bigg],
 \nonumber  \\
 g^{h_i}_{q_j\bar{q_4}}  &= r_{i1} \bigg[ (V^{q}_L)^{*}_{jj} (V^{q}_R)_{j4}  {\sqrt{2} m_{q_j} \over v }
						+ (V^{q}_L)^{*}_{j4} (V^{q}_R)_{4j}  { y_{j4}^{q} }
						+ (V^{q}_L)^{*}_{jj} (V^{q}_R)_{jj}  {(r_{i2} - i r_{i3}) \over r_{i1}} {\lambda_V^j} \bigg].
 \label{hqq_couplings}
 \end{align}

Moreover, such mixing will modify the interaction of quarks with the gauge bosons $V=Z,\,W^\pm$. 
The general form of quark couplings to $W^\pm$ for the left-handed (right-handed) sector may be given by
\begin{align}
&{g_{W}^{ij}}_{L/R} = {g \over \sqrt{2}} {(\delta_{L/R} V^{t}_{L/R}})^\dagger \cdot \delta_{L/R}  V^{b}_{L/R} = {g \over \sqrt{2}} (V_{L/R}^{CKMV})_{ij}, \label{WUD}
\end{align}
where $i,\, j = 1,\, \cdots, 4$ and the matrices $\delta_L=1_4$ and $\delta_R=diag(0,\,0,\,0,\, 1)$. In this model, the quark flavor mixing can be described with the CKMV matrix, where in the absence of vector quarks is reduced to the usual CKM matrix. 
This mixing would result in \textit{another} source of low-energy CP violation due to a constant phase in the CKMV matrix arising from the complex singlet's VEV~\cite{Enqvist,Branco:1985pf,Branco:1985ch,Branco:2003rt}. Therefore, we can parametrise the lowest weak-basis invariant in term of CKMV matrix elements and quark masses relevant to CP violation at $T=0$ as~\cite{Jarlskog,delAguila:1997vn}
\begin{align}
J_{L/R} = - m_{t}^2 M_{t'}^2 (M_{t'}^2 -m_{t}^2 )m_{b}^2 M_{b'}^2 (M_{b'}^2 -m_{b}^2 ) 
{\rm Im}(V^{tb}{V^{{t'b}^*}}V^{t'b'}{V^{{tb'}^*}})_{L/R},
\label{wbi}
\end{align}
that is required to satisfy the relation $n_B/s \approx 10^{-4} J / T_c^8 = (8.7 \pm 0.3) \times 10^{-11}$ for a successful baryogenesis. 

In addition, the coupling of quarks to $Z$ boson may be modified as
\begin{align}
&{g_{Z}^{ij}}_{\; L/R} = {g \over \cos\theta_W} \big(  T_3^i - 2 Q_i\sin^2\theta_W \big) \delta^{ij}+ \mathcal{S}^{t'(b')}_{L/R} {g \over \cos\theta_W}	(\delta_{L/R} V_{L(R)}^{t(b)})^{*}_{4i} (\delta_{L/R} V_{L/R}^{t(b)})_{4j}, \label{Zl}
\end{align}
where  $i,\, j$ run over all quarks and the weak isospin $T_3 = \pm1/2,0$. The second term is sensitive to the difference in isospin and may be parametrised as $\mathcal{S}^{t',b'}_{L}=0$, $\mathcal{S}^{t'}_{R}={1\over 2}$ and $\mathcal{S}^{b'}_{R}=-{1\over 2}$.
Consequently, due to the mixing between the new state and SM quarks (and between SM quarks themselves) a mixing is induced between quarks of different families. This rises to tree level FCNCs, which are absent in the SM at tree level and only can occur at loop-level.
However, there are constraints on FCNCs coming from a large number of observations that can therefore provide strong bounds on mixing parameters ~\cite{CMS:2017twu,Aaboud:2018nyl}. 
These limits can be translated into bounds on combinations of non-diagonal mixing matrix elements in both the left- and right-handed sectors. Moreover, direct bounds on heavy chiral quarks can be interpreted as bound on vector quarks, but decay channels of vector quarks are different from decay channels of heavy chiral quarks~\cite{Okada:2012gy}. Additionally, charged and neutral currents for vector quarks can have similar branching ratios, therefore with this
assumptions on the heavy-state decay channels, we may obtain the bounds on the heavy vector quark masses.
%The search for the FCNC top quark decays of BR$(t\to Zq)$ ($q$=$u$, $c$) in $pp$ collision at $\sqrt{s}=13$ TeV have been done both by CMS and ATLAS Collaborations through different channels ~\cite{CMS:2017twu,Aaboud:2018nyl}. 
%Upper limits for observed (expected) at 95$\%$ CL are set on the branching fractions of top quark decays BR$(t\to Zu)<  2.4 \times 10^{-4}(1.5 \times 10^{-4}$) and BR$(t\to Zc) < 4.5 \times 10^{-4}(3.7\times 10^{-4}$) from the CMS~\cite{CMS:2017twu}, BR$(t\to Zu) <  1.7\times 10^{-4}(2.4\times 10^{-4}$) and BR$(t\to Zc) < 2.4 \times 10^{-4} (3.2 \times 10^{-4}$) from the ATLAS~\cite{Aaboud:2018nyl} Collaborations.

\begin{figure}
\centering
\includegraphics[width=0.8\textwidth]{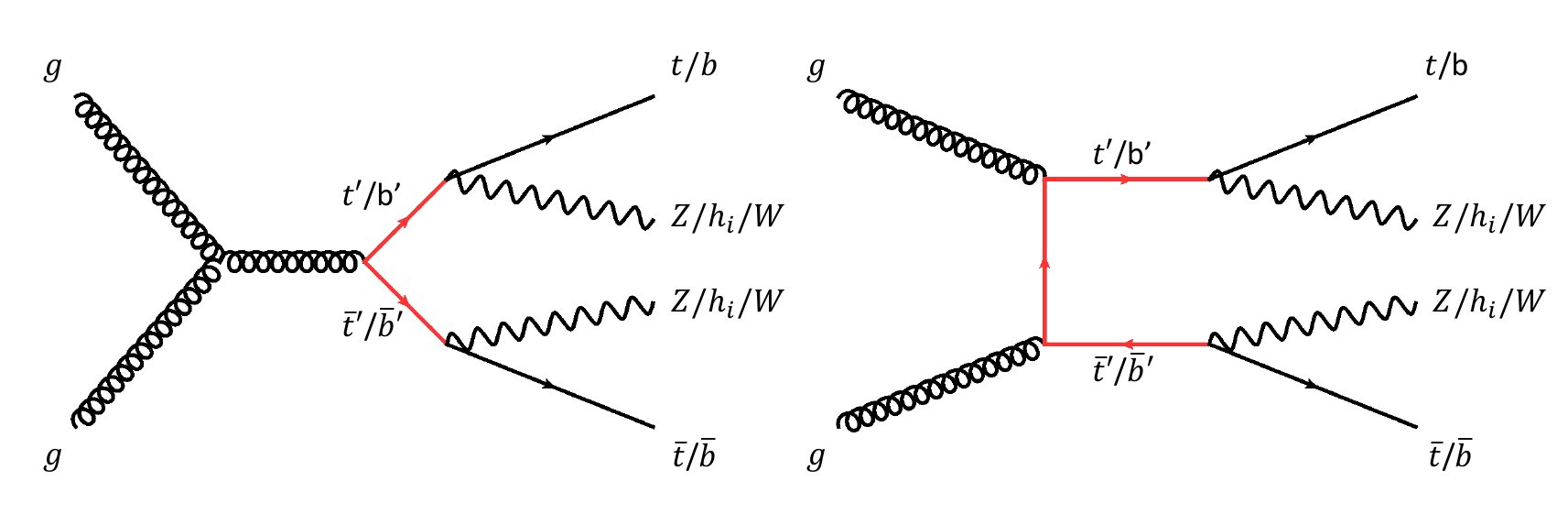}
\caption{ \it Dominant Feynman diagrams for $t'\bar{t'}$ and $b'\bar{b'}$ production and decay channels.}
\label{VQ}
\end{figure}

Here, we analyse the signatures of vector quarks at the LHC through decay channels $t' \to Wb,Zt,h_{i} t$ and $b' \to Wt,Zb,h_{i} b$.  The leading order Feynman diagrams for $b'\bar{b'}$ and $t'\bar{t'}$ productions and decays are shown in Figure~\ref{VQ}.

The partial widths for $t' \to Wb,Zt,h_{i}t$ and  $b' \to Wt,Zb,h_{i} b$ channels are given by
\begin{eqnarray}
&&\Gamma\big( t' \to Wb, Zt, h_i t \big) = \kappa^2_{W,Z,h_i} \big|V^{t'q}_{L/R} \big|^2
	{M_{t'}^3 g^2\over 64\pi M_{W}^2} \times \Gamma_{W,Z,h_i}\big( M_{t'}, M_{W,Z,h_i}, m_{q} \big), \\
&&\Gamma\big( b' \to Wt, Zb, h_i b \big) = \kappa^2_{W,Z,h_i} \big|V^{b'q}_{L/R} \big|^2
	{M_{b'}^3 g^2 \over 64\pi M_{W}^2} \times \Gamma_{W,Z,h_i}\big( M_{b'}, M_{W,Z,h_i}, m_{q} \big),
	\label{gamma-VQ}
\end{eqnarray}
where $V^{t'q}_{L/R}$ and  $V^{b'q}_{L/R}$ represent the $L/R$
mixing matrices between the new quarks and the SM quarks, labelled by $q$, while the parameters $\kappa_{W,Z,h_i} $ are normalized couplings to the gauge bosons and Higgs particles. 
The kinematic relations for $\Gamma_{W,Z}$ and $\Gamma_{h_i}$ can be given by
\begin{eqnarray}
&&\Gamma_{W,Z}={1 \over 2} \sqrt{F\bigg( 1, {m_q^2 \over M_{t',b'}^2}, 
	{M_{W,Z}^2 \over M_{t',b'}^2} \bigg)} \Bigg[ \bigg( 1- {m_q^2 \over M_{t',b'}^2}\bigg)^2 
	+ {M_{W,Z}^2 \over M_{t',b'}^2} 
	- 2 {M_{W,Z}^4 \over M_{t',b'}^4} 
	+ {m_{q}^2 M_{W,Z}^2 \over M_{t',b'}^4} \Bigg] ,	
\\
&&\Gamma_{h_i}={1 \over 2} \sqrt{F\bigg( 1, {m_q^2 \over M_{t',b'}^2}, 
	{M_{h_i}^2 \over M_{t',b'}^2} \bigg)} \Bigg[ 
	1 + {m_{q}^2 \over M_{t',b'}^2} 
	- {M_{h_i}^2 \over M_{t',b'}^2}  \Bigg],
\end{eqnarray}
where the function $F(a, b, c)$ is
$$
F(a, b, c)=a^2+b^2+c^2-2ab-2ac-2bc.
$$
Neglecting quark masses, $m_q=0$, the branching rations can be written as
\begin{eqnarray}
&&\text{BR}(t' \to V q) = {\kappa_{V}^2 \big| V^{t'q}_{L/R} \big|^2 \Gamma^0_{V} 
	\over 
	\big( \sum_{q=1}^3 \big| V^{t'q}_{L/R} \big|^2 \big)
	\big( \sum_{V'=W,Z,h_i} \kappa_{V'}^2 \Gamma^0_{V'} \big)
	} 	,
	\\
&& \text{BR}(b' \to V q) = {\kappa_{V}^2 \big| V^{b'q}_{L/R} \big|^2 \Gamma^0_{V} 
	\over 
	\big(\sum_{q=1}^3 \big| V^{b'q}_{L/R} \big|^2 \big)
	\big( \sum_{V'=W,Z,h_i} \kappa_{V'}^2 \Gamma^0_{V'} \big)
	},
\end{eqnarray}
where the branching fractions translate directly to the squares of the magnitudes of the corresponding $V^{q'q}_{L/R}$ elements.
In the limit of massless particles, the $\Gamma^0_{Z} $, $\Gamma^0_{W} $ and $\Gamma^0_{h_i}$ may be obtained as
%% ----------------------------------------------------------------- %%
\begin{align}
&\Gamma_{W}^0= \bigg( 1 - 3{M_{W}^4 \over M_{t',b'}^4} + 2{M_{W}^6 \over M_{t',b'}^6} \bigg) 
	\sim 1 + \mathcal{O}(M_{t',b'}^{-4}),
\\
&\Gamma_{Z}^0= {1 \over 2} \bigg( 1 - 3{M_{Z}^4 \over M_{t',b'}^4} + 2{M_{Z}^6 \over M_{t',b'}^6} \bigg) 
	\sim {1 \over 2} + \mathcal{O}(M_{t',b'}^{-4}),
\\
&\Gamma_{h_i}^0= {1 \over 2} \bigg( 1 - {M_{h_i}^2 \over M_{t',b'}^2} \bigg)^2 
	\sim {1  \over 2} - {M_{h_i}^2 \over M_{t',b'}^2} + \mathcal{O}(M_{t',b'}^{-4}).
\end{align}
%% ----------------------------------------------------------------- %%
  In the most general set-up, $t'$ and $b'$ may have sizeable couplings to both left- and right-handed chiral quarks. However, only one of the two mixing angles is large and the other are suppressed by a factor of $m_q/M$. Following this observation, we can simplify the parametrisation by neglecting the suppressed mixing angles.

\begin{figure*}[t]
\includegraphics[width=1\textwidth]{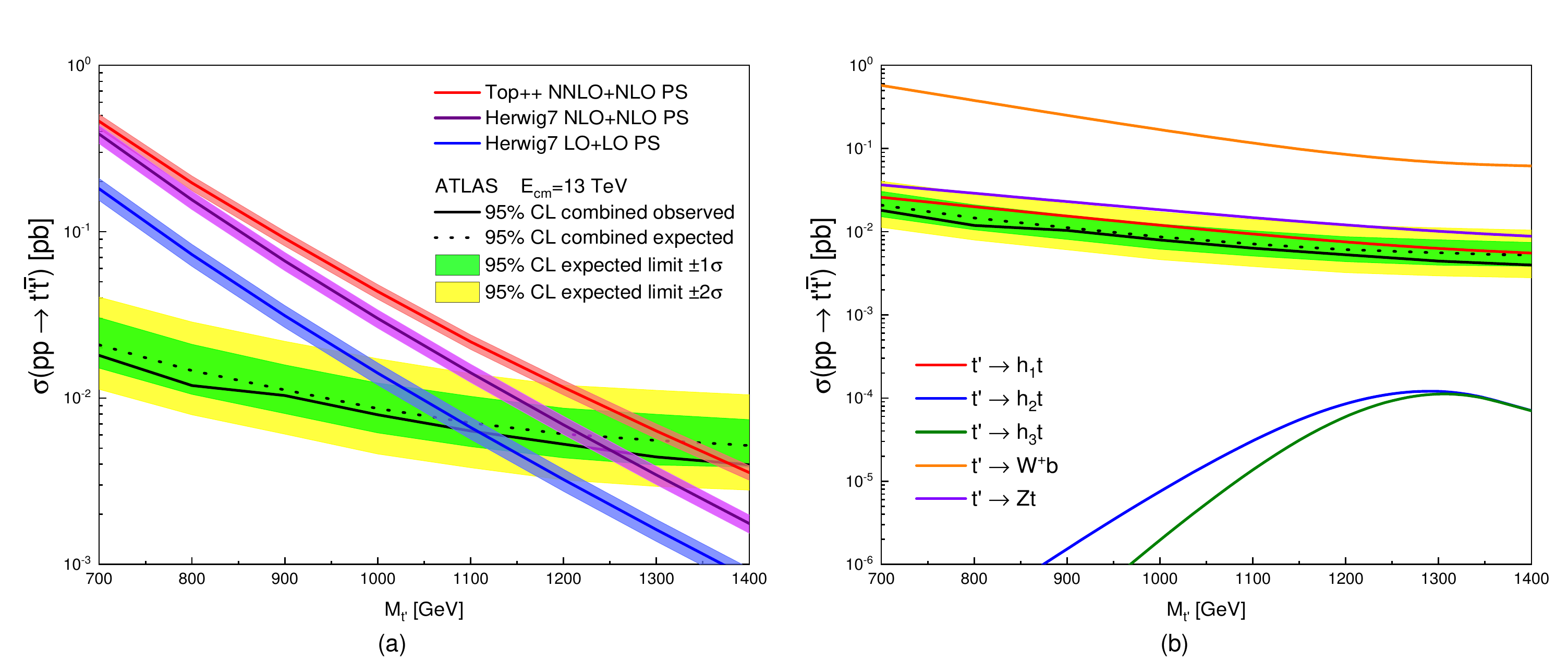}
\caption[]{\it The $t' \bar{t'}$  pair production cross-section is plotted versus $M_{t'}$. (a) The LO and NLO predictions are generated by \textsf{MadGraph5} \cite{Alwall:2014hca}, and enhanced with LO and NLO PSs from \textsf{Herwig7} \cite{Bahr:2008pv,Bellm:2019zci}. The NNLO results are from~\cite{Aaboud:2018pii}, computed with the LO generator \textsf{Protos}~\cite{PROTOS}, showered with \textsf{Pythia8}~\cite{Sjostrand:2007gs} and normalized with \textsf{Top++}~\cite{Czakon:2011xx}. These results are compared with the corresponding upper limits prediction from ATLAS~\cite{Aaboud:2018pii}. (b) The individual $t' \to Wb,Zt,h_{i}t$ and  $b' \to Wt,Zb,h_{i} b$ decay channels are displayed. The properties of Higgs boson $h_1,\, h_2$ and $h_3$ are taken from benchmark $A1$ in Section~\ref{sec:4}.
} 
\label{figMtp}
\end{figure*}

\begin{figure*}[t]
\includegraphics[width=1\textwidth]{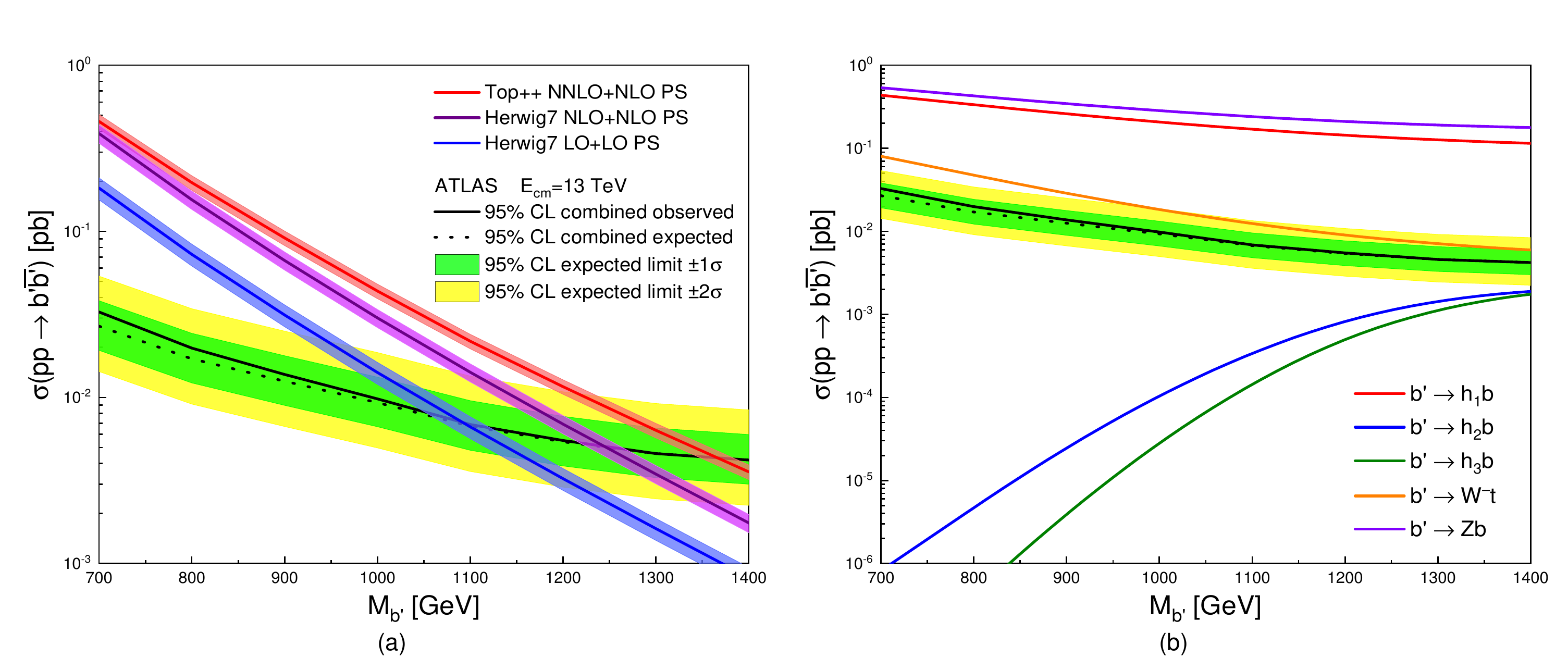}
\caption[]{\it The same as in Figure~\ref{figMtp}, the $b' \bar{b'}$ pair production cross-section is plotted versus  $M_{b'}$. } 
\label{figMbp}
\end{figure*}

In Figures \ref{figMtp} and \ref{figMbp}, we predict the total cross-section for $pp \to t'\bar{t'}$ and $pp \to b'\bar{b'}$ production channels in the cSMCS model and compare our results with the corresponding upper limits prediction from ATLAS to obtain the bounds on the heavy quark masses $M_{t'}$ and $M_{b'}$. 
The observed (solid black line) and the expected (dotted line) 95$\%$ CL upper limits are from the ATLAS collaboration predictions~\cite{Aaboud:2018pii}. The shaded bands correspond to $\pm 1\sigma$ and $\pm 2\sigma$ standard deviations around the combined expected limit. These are a combination of the searches for pair-produced vector quarks $t'\bar{t'}$ and $b'\bar{b'}$ in various decay channels $t' \to Wb,Zt,H_{\rm SM} t$ and  $b' \to Wt,Zb,H_{\rm SM} b$ performed using 36.1 fb$^{-1}$ of $pp$ collision data at $\sqrt{s} = 13$ TeV with the ATLAS detector at the LHC.

In the left panels of Figures \ref{figMtp} and \ref{figMbp}, our predictions for $\sigma(pp \to t'\bar{t'})$ and $\sigma(pp \to b'\bar{b'})$ are presented with leading-order (LO), next-to-leading-order (NLO) and next-to-next-to-leading-order (NNLO) accuracies within their corresponding uncertainty bounds. The LO and NLO predictions are generated by \textsf{MadGraph5} \cite{Alwall:2014hca}, and enhanced with LO and NLO parton showers (PSs) from \textsf{Herwig7} \cite{Bahr:2008pv,Bellm:2019zci}. The NNLO calculations are from~\cite{Aaboud:2018pii}, computed with the LO generator \textsf{Protos}~\cite{PROTOS}, showered with \textsf{Pythia8}~\cite{Sjostrand:2007gs} and normalized with \textsf{Top++}~\cite{Czakon:2011xx} at NNLO~QCD accuracy. Accordingly, we obtain the lower bounds on the heavy quark masses to be $M_{t'}\sim M_{b'}\gtrsim 1.4$~TeV.

Furthermore, in the right panels of Figures \ref{figMtp} and \ref{figMbp}, the individual $t' \to Wb,Zt,h_{i}t$ and  $b' \to Wt,Zb,h_{i} b$ decay channels are plotted and compared with the corresponding 95$\%$~CL limits from ATLAS. Additionally, the topologies of $b'$ and $t'$ decays can be determined by the following branching fractions, 
\begin{align}
& {\rm BR}(t' \to Z t)+{\rm BR}(t' \to W^- b)+\sum_{i=1}^3 {\rm BR}(t' \to h_i t) =1,
 \\
& {\rm BR}(b' \to Z b)+{\rm BR}(b' \to W^+ t)+\sum_{i=1}^3 {\rm BR}(b' \to h_i b) =1.
\end{align}
consequently, we may obtain the following bounds 
\begin{align}
& {\rm BR}(t' \to Z t) + \sum_{i=1}^3 {\rm BR}(t' \to h_i t)  \sim  1,
\\
& {\rm BR}(t' \to W^- b) \ll 0.1, \\
& {\rm BR}(b' \to Z b)+\sum_{i=1}^3 {\rm BR}(b' \to h_i b)  \ll 0.1,
\\
& {\rm BR}(b' \to W^+ t) \sim  1.
\end{align}
Hence, one can readily identify the decay channels $t' \to Zt$, $t' \to h_{1}t$ and $b' \to W^+ t$ to be the most promising probes in a search for the heavy vector quarks within the cSMCS model. 

Additionally, having obtained the bound on the masses of vector quarks and branching fractions $\text{BR}(q' \to V q) = {|V^{q'q}}|^2$, we may estimate a very small ${\rm BR}(t' \to V b')$ to fulfill the constraint on the Jarlskog invariant, Eq. \eqref{wbi}, for a successful baryogenesis. Furthermore, the branching fraction ${\rm BR}(t' \to V b')$ can be controlled by the mass-splitting $M_{t'}-M_{b'}$ that should be within the acceptable boundaries of the oblique parameters~\cite{Kribs:2007nz}, to be discussed in the next section. 

 
 In the next section, we will present our numerical estimates of the SM-like Higgs boson signal strength, $\mathcal{R}_{\gamma\gamma}$, as well as corrections that would arise to the gauge boson propagators, considering $M_{t'} \sim M_{b'} = 1.45$ TeV from the limit given in Figures \ref{figMtp} and \ref{figMbp}. 
 
\section{Contribution of vector quarks to Higgs production and decay}
\label{sec:4}

As we have seen in Section \ref{sec:3}, the cSMCS model allows for a SM-like Higgs particle observed at the LHC. Compared with the SM in the absence of vector quarks, the processes 
$H_{\rm SM} \to VV^{*}$ ($V = W^{\pm}, Z$) and  $H_{\rm SM} \to gg$ and $H_{\rm SM} \to \gamma \gamma$ are only suppressed by the factor~$r_{11}^2$.
This is since, the SM-like Higgs field in this model can be identified by the following linear field combination,
\begin{align}
h_1 &= c_1 c_2 \phi_1 + (c_3 s_1 - c_1 s_2 s_3) \phi_2 + (c_1 c_3 s_2 + s_1 s_3) \phi_3
\nonumber \\
&=r_{11} \phi_1 +r_{12} \phi_2 + r_{13} \phi_3,
\end{align}
where both $r_{12}$ and $r_{13}$ are negligible.
Henceforth, for example the SM-normalised partial decay width of the lightest Higgs boson $h_1$ to the EW gauge bosons ($V = W^{\pm}, Z$) takes on the form,
 \begin{align}
 \Gamma(h_{1} \to VV^{*})\approx r_{11}^{2} \Gamma(H_{\rm SM} \to VV^{*} ),
 \end{align}
that remains unchanged with the addition of vector quarks.

However, as we have shown in Figure~\ref{VQ}, these heavy vector quarks contribute in the amplitudes for Higgs production via $gg \to t'\bar{t'}/b'\bar{b'} \to h_{1} X$ channels as well as Higgs decay into photons. These extra contributions would directly effect the Higgs boson signal strength, 
\begin{align}
\label{Rxx}
 \mathcal{R}_{\gamma \gamma} =& \frac{ \sigma( pp\to h_1 X) }{ \sigma( pp\to H_{\rm SM} X) } 
 \frac{\text{BR}(h_1 \to \gamma \gamma) }{\text{BR} (H_{\rm SM} \to\gamma \gamma) }.
\end{align}
Now, by assuming the gluon fusion as the dominant Higgs production channel at the LHC and the narrow-width approximation, the expression for $\mathcal{R}_{\gamma \gamma}$ may reduce to
\begin{align}
\label{Rxx_2}
 \mathcal{R}_{\gamma \gamma} =& \frac{ \Gamma(h_1 \to gg) }{ \Gamma(H_{\rm SM} \to gg)} 
 \frac{\Gamma(h_1 \to \gamma \gamma)}{\Gamma(H_{\rm SM} \to \gamma \gamma)}
  \frac{\Gamma^{H_{\rm SM}}_{\rm tot}}{\Gamma^{h_1}_{\rm tot}},
\end{align}
where the total decay width of the lightest Higgs boson is given as a combination of the following partial decay widths
\begin{align}
\Gamma_{\rm tot}^{h_1} =  \sum_{f=l,q} \Gamma(h_1 \to f\bar{f}) 
						+ \sum_{V=W^{\pm},Z,\gamma,g} \Gamma(h_1 \to VV^{*})
						+ \Gamma(h_1 \to Z\gamma).
\label{total_decay_width}
\end{align}
In the Eqs.~\eqref{Rxx_2} and \eqref{total_decay_width}, the relevant one-loop partial decay widths of the lightest Higgs boson are given by
\begin{align}
\Gamma(h_1 \to \gamma \gamma) &= {\alpha_{\rm em} \alpha^2 M_{h_1}^3 \over 256 \pi^2 M_W^2} 
	\bigg| 
	\sum_{q=t,b,t',b'} Q_q^2 N_c \; \kappa^{qq}_{h_1} \; A_{1/2}(x_q^{h_1}) + r_{11} A_{1}(x_W^{h_1})
	\bigg|^2,
	\label{h2gaga}
	\\
	\Gamma(h_1 \to gg) &= {\alpha_{\rm em} \alpha_{\rm s}^2 M_{h_1}^3 \over 64 \pi^2 M_W^2} \bigg| 
	\sum_{q=t,b,t',b'} Q_q^2 N_c \; \kappa^{qq}_{h_1} \; A_{1/2}(x_q^{h_1})
	\bigg|^2 ,
	\label{h2gg}
	\\	
	\Gamma(h_1 \to Z \gamma) &= {\alpha_{\rm em}^2 \alpha M_{h_1}^3 \over 128 \pi^2 M_W^2} \bigg( 1- {M_Z^2 \over M_{h_1}^2} \bigg)^3 
	\bigg| 
	\sum_{q=t,b,t',b'} 2 Q_q N_c \; \kappa^{qq}_{Z} \; \kappa^{qq}_{h_1} B_{1/2}(x_q^{h_1},y_q) 	 
	\nonumber \\	
	& + r_{11} A_{1}(x_W^{h_1}) 
	+ \sum_{q,q'=t,b,t',b'} \sum_{c=L,R} {m_q \over M_W} (\kappa^{qq'}_{Z})_{c} \;
	\kappa^{qq'}_{h_1} \; B_{1/2}(x_q^{h_1},y_{q'})
	\bigg|^2,
\label{h2zga}
\end{align}
with $N_c=3$, $x_q^{h_1} = 4m_{q}^2 / M_{h_1}^2$, $x_W^{h_1} = 4M_{W}^2 / M_{h_1}^2 $ and $y_q = 4 m_{q}^2 / M_{Z}^2 $. In the above equations, $\kappa_{h_1,Z}$ are the couplings~\eqref{hqq_couplings} and \eqref{Zl} that are normalized by the SM couplings $y_t$ and $g$. Also, the loop functions $A_s(x)$ and $B_s(x,y)$ are given in the Appendix \ref{appA}. 

Note that, the total decay width for the heavier Higgs bosons would be significantly modified, if they can decay into the lighter Higgs bosons. In this case, the total decay width may be written as
\begin{align}
\Gamma_{\rm tot}^{h_i} &=  \sum_{f=l,q} \Gamma(h_i \to f\bar{f}) 
						+ \sum_{V=W^{\pm},Z,\gamma,g} \Gamma(h_i \to VV^{*})
						\nonumber \\
						&+ \sum_{j<i} \Gamma(h_i\to h_j h_j)
						+ \Gamma(h_i \to Z\gamma),
\label{total_decay_width_i}
\end{align}
where the partial decay width for $h_i \to h_j h_j$ ( $i>j$) is
\begin{equation}
 \Gamma(h_i\to h_j h_j)=\frac{g_{h_i h_j h_j}^{2}}{32\pi M_{h_{i}}}\left(1-\frac{4M_{ h_j}^{2}}{M_{h_{i}}^{2}}\right)^{1/2}.\label{invdec} 
 \end{equation}
The couplings $g_{h_i h_j h_j}$ are given in Appendix~\ref{hhh}.

Hence, combining the above equations we may calculate $\mathcal{R}_{\gamma\gamma}^{h_i}$, which is enhanced due to the contributions of the heavy vector quarks in our model with respect to its value in the absence of heavy vector quarks, ${\mathcal{R}^{h_i \; 0}_{\gamma\gamma}} \approx r_{i1}^2 \lesssim 1$.

\begin{figure*}[t]
\includegraphics[width=0.55\textwidth]{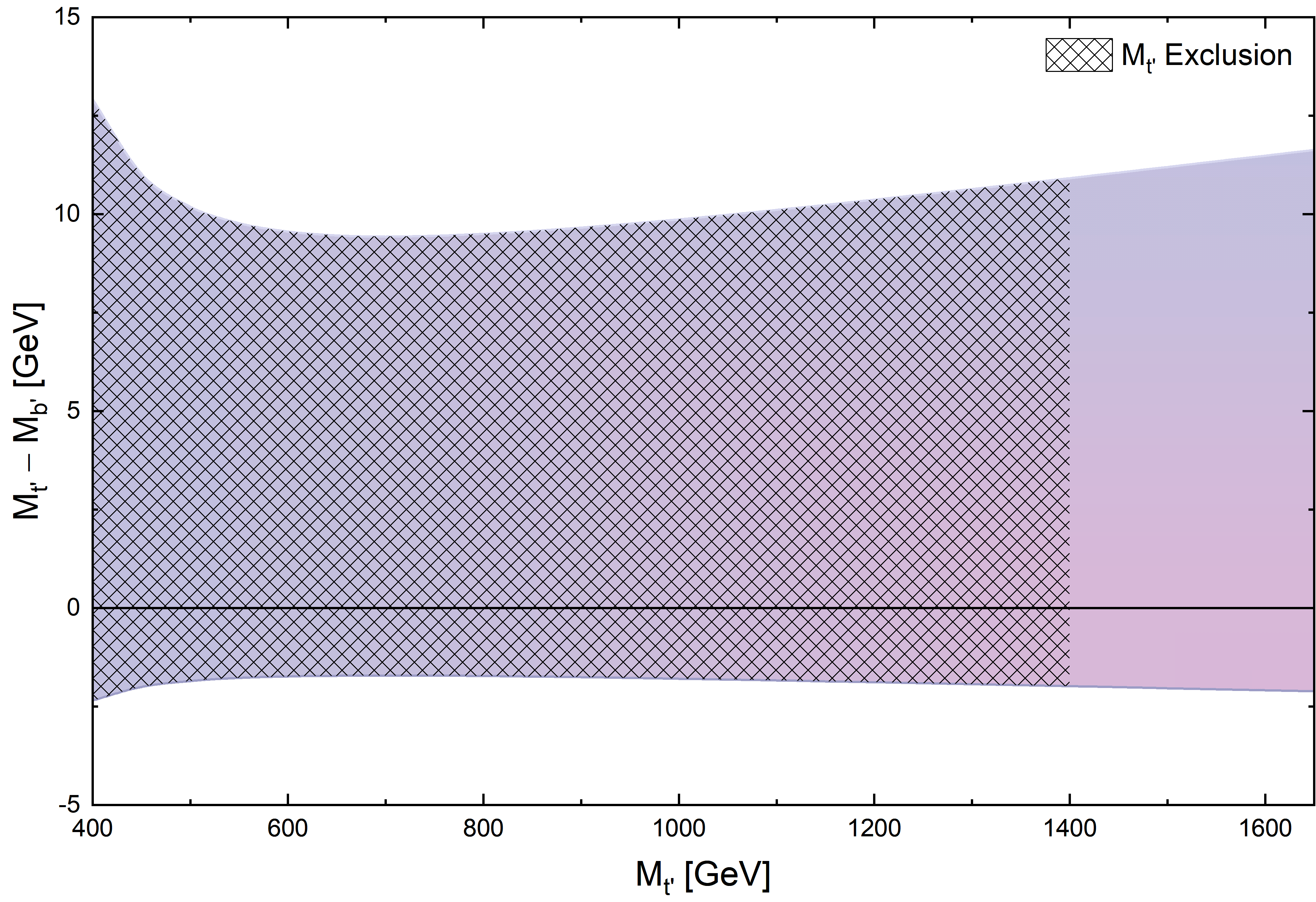}
\caption[]{\it Allowed range for the mass splitting of the heavy quark. } 
\label{mstb}
\end{figure*}

Furthermore, because of the presence of additional particles in the model, some corrections to the gauge boson propagators would arise. These corrections can be parametrised by the oblique parameters $S$ and $T$ ~\cite{Peskin:1991sw}. In the limit $m_b \to 0 $, the contributions of the new heavy quarks into  $S$ and $T$ parameters are well approximated by \cite{Dawson:2012di},
\begin{align}
& \Delta T_{t',b'} \approx {12 \over 16 \pi \sin^2 \theta_w} {m_t^2 \over M_{W}^2} {(M_{t'}-M_{b'})\over M_{t'}} \Big[2 \log{r} -3+ {5 \log{ r} -3 \over r}  \Big], \\ 
& \Delta S_{t',b'} \approx {1 \over 3 \pi } {(M_{t'}-M_{b'})\over M_{t'}} \Big[4 \log{r} -7+ {4 \log{ r} -7 \over r}  \Big],
\end{align}
where $r=M_{t'}^2/ m_t^2$.  In Figure~\ref{mstb}, we plot the range of mass splitting $M_{t'}-M_{b'}$ versus $M_{t'}$ within the acceptable boundary for $S = 0.05\pm0.11$ and $T = 0.09\pm0.13$~\cite{Baak:2014ora}. The coloured area above $1.4$ TeV indicates the allowed region of the theoretical predictions and the experimental bounds on $M_{t'}$, as shown in Figure~\ref{figMtp}.
From Figure~\ref{mstb}, we observe that a small mass splitting between vector quarks $-2\lesssim M_{t'}-M_{b'} \lesssim 10$~GeV is required to be consistent with the precision electroweak measurement. This leads to the suppression of cascade decays such as $ t' \to W b'$.
Other modifications to the oblique parameters in this model are induced by the presence of the new Higgs particles~\cite{Chen:2003fm,Grimus:2008nb}. These would allow the small mass-splitting of the vector quarks to have larger values, which is important in the parametrisation of the lowest weak-basis invariant for the low-energy CP violation. The oblique parameters $S$ and $T$ arise from the new Higgs bosons are parametrised in~Appendix~\ref{A-obli}.

In Table \ref{tab-1},  we propose 5 benchmarks containing the masses for Higgs bosons $h_{1}$, $h_{2}$ and $h_{3}$ and their mixing angles $\alpha_{1}$, $\alpha_{2}$ and $\alpha_{3}$.
This table also contains our predictions for the $S$ and $T$ parameters that are induced by the new scalars and the heavy vector quarks with mass splitting $M_{t'}- M_{b'}\sim 1$~GeV and $M_{t'}=1.45$~TeV. These results are consistent with the current data within $3 \sigma$ uncertainty level.
In Table \ref{tab-2}, we compare our predictions for $\mathcal{R}_{\gamma \gamma}$ with/without the contributions of the heavy vector quarks and considering $M_{t'}\sim M_{b'}$. 
Notice that, these additional contributions from the heavy vector quarks would significantly enhance the predicted values of $\mathcal{R}_{\gamma \gamma}$.
Moreover, we confront our predictions with
existing experimental data from ATLAS and CMS, including their
statistical and systematic uncertainties~\cite{Aad:2014eha,Khachatryan:2014ira}. 
Our predictions for $\mathcal{R}_{\gamma \gamma}$ are in excellent agreement with the observed results. 

\begin{table*}[t]
\begin{tabular}{|c|c|c|c|c|c|c|c|c|c|c|c|}
\hline
 Benchmark & $\alpha_1$ &$\alpha_2$&$\alpha_3$&$M_{h_{1}}$ &$M_{h_{2}}$&$M_{h_{3}}$&$S$&$T$ \\ \hline
 
 $A1$ &-0.047 &-0.053 &1.294 & 124.64 & 652.375 & 759.984 & -0.071 & -0.074 \\ \hline 
 $A2$ &-0.048 &0.084 &0.084 & 124.26 & 512.511 & 712.407 & -0.001 & -0.019 \\ \hline 
 $A3$ &0.078 &0.297 &0.364 & 124.27 & 582.895 & 650.531 & 0.003 & -0.024  \\ \hline 
 $A4$ &0.006 &-0.276 &0.188 & 125.86 & 466.439 & 568.059 & -0.012 & -0.149  \\ \hline
 $A5$ &0.062 &-0.436 &0.808 & 125.21 & 303.545 & 582.496 & 0.002 & -0.389  \\ \hline
\end{tabular}
\caption{\it Benchmark points $A1-A5$ show mixing angels $\alpha_{1,2,3}$, Higgs masses $M_{h_{1,2,3}}$ and Oblique parameters $S$ and $T$. The contributions to the Oblique parameters include both the new scalars and the heavy vector quarks with mass splitting $M_{t'}- M_{b'}\sim 1$~GeV and $M_{t'}=1.45$~TeV. The values that are determined from a fit with reference mass-values of top and Higgs boson $M_{t,ref}=173 \; {\rm GeV}$ and $M_{h,ref}=125 \; {\rm GeV} $ read; 
$S = 0.05\pm0.11,\, T = 0.09\pm0.13$~\cite{Baak:2014ora}.}\label{tab-1}
\vspace{.2in}
\begin{tabular}{|c|c|c|c|c|c|c|c|c|c|c|c|}
\hline
 Benchmark &${\mathcal{R}^{h_1 \; 0}_{\gamma\gamma}}$ &$ \mathcal{R}^{h_1}_{\gamma \gamma}$ & ${\mathcal{R}^{h_2 \; 0}_{\gamma\gamma}}$&$ \mathcal{R}_{\gamma\gamma}^{h_{2}}$& ${\mathcal{R}^{h_3 \; 0}_{\gamma\gamma}}$&$ \mathcal{R}_{\gamma\gamma}^{h_{3}}$\\ \hline
 
 $A1$ &0.9800 &1.1322 &0.0021 &0.0045 &0.0028 & 0.0018 \\ \hline 
 $A2$ &0.9800 &1.1272 &0.0021 &0.0026 &0.0070 & 0.0020 \\ \hline 
 $A3$ &0.9800 &1.0340 &0.0055 &0.0077 &0.0850 & 0.1928 \\ \hline 
 $A4$ &0.9200 &1.0533 &0.0000 &0.0000 &0.0740 & 0.1476 \\ \hline
 $A5$ &0.8100 &0.9317 &0.0029 &0.0035 &0.1700 & 0.2371 \\ \hline
\end{tabular}
\caption{\it Values of Higgs signal strength ${\mathcal{R}^{h_i}_{\gamma\gamma}}$ ($ \mathcal{R}^{h_i \; 0}_{\gamma \gamma}$) with (without) the contributions of the heavy vector quarks, considering $y_{34} \sim 0.3$, $\lambda_V \sim 0.01$ and $M_{t'}\sim M_{b'} =1.45$~TeV for benchmark points $A1-A5$. The best-fit signal strength $\mathcal{R}_{\gamma\gamma}$ relative to the SM prediction from ATLAS,  CMS and the combined analysis ATLAS+CMS within $1\sigma$ deviation are $\mathcal{R}_{\gamma\gamma}^{}=1.14^{+0.27}_{-0.25}$~\cite{Aad:2014eha},  ~$\mathcal{R}_{\gamma\gamma}^{}=1.11^{+0.26}_{-0.23}$~\cite{Khachatryan:2014ira} and ~$\mathcal{R}_{\gamma\gamma}^{}=1.14^{+0.19}_{-0.18}$~\cite{201606}, respectively.}\label{tab-2}
\end{table*}

Moreover, we have justified the validity of the above benchmarks with respect to the existing experimental bounds using the \texttt{HiggsBounds} package (version 5.3.2)~\cite{Bechtle:2008jh,Bechtle:2015pma}. This includes constraints from direct Higgs searches at the LEP, the Tevatron and the LHC with the most sensitive exclusion limit for each parameter at 95$\%$ confidence interval. 

\section{Conclusions }\label{sec:con}

We have investigated one of the simplest extensions of the SM with a complex singlet and a pair of heavy doublet vector quarks, the so-called cSMCS model.
The potential of this model contains 13 real parameters which can be simplified by an accidental symmetry of the model. Here, we have considered a global U(1)-symmetry with some U(1)-soft breaking terms.
In this model, the CP violation can emerge spontaneously as a consequence of the time-dependent phase of the complex singlet VEV, which gives rise in the mass mixing of the SM and heavy vector quarks. We have shown that the time-dependent CP-violating phase depends on the Higgs field within the bubble-wall via the Higgs-singlet coupling that can directly explain the observed baryon-to-entropy ratio $\sim 9\times 10^{-11}$. 
Moreover, we have shown the parameter domains in the cSMCS model for the non-vanishing phase and for the limit allowed by SM-like Higgs boson mass. 

Furthermore, the mixing between vector quarks and SM quarks 
giving rise to tree-level FCNC and another source of CP violation at low temperature via  $t' \to Wb,Zt,h_{i} t$ and $b' \to Wt,Zb,h_{i} b$ decay channels. However, there are constraints on these channels from a large number of observations that can therefore provide strong bounds on mixing parameters ~\cite{CMS:2017twu,Aaboud:2018nyl}.  We have investigated the implications of these constraints on the total cross-section of $pp \to t'\bar{t'}$ and $pp \to b'\bar{b'}$ production channels and have obtained the lower bounds on the heavy quark masses to be $M_{t'}\sim M_{b'}\gtrsim 1.4$~TeV. New heavy vector quarks contribute in the amplitudes of Higgs via $pp \to t'\bar{t'}/b'\bar{b'}\to h_i X$ production and $h_i \to \gamma \gamma$ decay channels. Accordingly, we have shown the contributions of the vector quarks in the Higgs bosons signal strength, $\mathcal{R}_{\gamma\gamma}$, as well as corrections to the gauge boson propagators. Additionally, our framework is consistent with the existing experimental bounds from direct Higgs searches at the LEP, the Tevatron and the LHC with the most sensitive exclusion limit for each parameter at 95$\%$ CL. 

Finally, our benchmarks have been used to predict the production rates of the Higgs bosons at the LHC, where the lightest Higgs production cross-section can be fitted to the experimental data from ATLAS and CMS at the $2\sigma$ level~\cite{Darvishi:2016fwo}. These predictions for yet undetected heavy Higgs bosons and heavy quarks of the cSMCS model may provide some clues for the future discovery.

\section*{Acknowledgements}
We are grateful to G. Branco, I. Ivanov, M.R. Masouminia, A. Pilaftsis, M. Rebelo and  M. Sampaio for constructive discussions. We also express our special thanks to D. Soko\l owska for valuable suggestions and S. Najjari for support. This work was partially supported by the grant NCN OPUS 2012/05/B/ST2/03306 (2012-2016).
The work of ND is also supported in part by the Lancaster—Manchester—Sheffield
Consortium for Fundamental Physics, under STFC research grant ST/P000800/1 and the National Science Centre, Poland, the HARMONIA project under contract UMO-2015/18/M/ST2/00518 (2016-2021).

%%%%%%%%%%%%%%%%%%%%%%%%%%%%%%%%%%%%%%%%%%%%%%%%%%%%%%%%%%%%%%%%%%%%%%%%%%%%%%%%%%%%%%%%%%%%%%%%%%%%%%%%%%%%%%%%%%%%%%%%%%%%%%%%%%%
\appendix
\section{One-loop Higgs decay widths}\label{appA}

In Section \ref{sec:4}, we have shown that vector quarks would contribute in the one-loop Higgs bosons partial decay widths $\Gamma(h_i \to gg)$, $\Gamma(h_i \to \gamma \gamma)$ and $\Gamma(h_i \to Z \gamma)$, 
\begin{align}
\Gamma(h_i \to \gamma \gamma) &= {\alpha_{\rm em} \alpha^2 M_{h_i}^3 \over 256 \pi^2 M_W^2} 
	\bigg| 
	\sum_{q=t,b,t',b'} Q_q^2 N_c \; \kappa^{qq}_{h_i} \; A_{1/2}(x_q^{h_i}) + r_{i1} A_{1}(x_W^{h_i})
	\bigg|^2,
	\\
	\Gamma(h_i \to gg) &= {\alpha_{\rm em} \alpha_{\rm s}^2 M_{h_i}^3 \over 64 \pi^2 M_W^2} \bigg| 
	\sum_{q=t,b,t',b'} Q_q^2 N_c \; \kappa^{qq}_{h_i} \; A_{1/2}(x_q^{h_i})
	\bigg|^2 ,
	\\	
	\Gamma(h_i \to Z \gamma) &= {\alpha_{\rm em}^2 \alpha M_{h_i}^3 \over 128 \pi^2 M_W^2} \bigg( 1- {M_Z^2 \over M_{h_i}^2} \bigg)^3 
	\bigg| 
	\sum_{q=t,b,t',b'} 2 Q_q N_c \; \kappa^{qq}_{Z} \; \kappa^{qq}_{h_i} B_{1/2}(x_q^{h_i},y_q) 	 
	\nonumber \\	
	& + r_{i1} A_{1}(x_W^{h_i}) 
	+ \sum_{q,q'=t,b,t',b'} \sum_{c=L,R} {m_q \over M_W} (\kappa^{qq'}_{Z})_{c} \;
	\kappa^{qq'}_{h_i} \; B_{1/2}(x_q^{h_i},y_{q'})
	\bigg|^2,
\label{h2gaga-Ap}
\end{align}
where $x_q^{h_i} = 4m_{q}^2 / M_{h_i}^2$, $x_W^{h_i} = 4M_{W}^2 / M_{h_i}^2 $ and $y_q = 4 m_{q}^2 / M_{Z}^2 $.
The loop functions for the above relations are given by
\begin{align}
A_{1}(x) &= -\left[2 + 3x + 3x(2 - x)f(x)\right], \\
A_{1/2}(x) &= 2 x \left[1+(1-x)f(x)\right], \\
B_{1}(x,y)&=\cos\theta_{W} \Bigg\{ 4\left(3-\frac{\sin^{2}\theta_{W}}{\cos^{2}\theta_{W}}\right)I_{2}(x,y)
\nonumber \\
&+\left[\left(1+\frac{2}{x}\right)\frac{\sin^{2}\theta_{W}}{\cos^{2}\theta_{W}}-\left(5+\frac{2}{x}\right)\right]I_{1}(x,y) \Bigg\},
\\
B_{1/2}(x,y)&=I_{1}(x,y)-I_{2}(x,y), 
\end{align}
where
\begin{align}
f(x) &= \left\{
\begin{matrix}
\arcsin^2(1/\sqrt{x}) & \;\;\;\; &  x \geq 1 \\
-{1\over 4}\bigg[ \log {1+\sqrt{1-x} \over 1-\sqrt{1-x}} -i\pi \bigg]^2 & \;\;\;\; & x < 1 
\end{matrix}
\right. ,
\\
I_{1}(x,y)&=\frac{xy}{2(x-y)}+\frac{x^{2}y^{2}}{2(x-y)^{2}}\left[f(x)-f(y)\right]+\frac{x^{2}y}{(x-y)^{2}}\left[g(x^{-1})-g(y^{-1})\right],
\\
I_{2}(x,y)&=-\frac{xy}{2(x-y)}\left[f(x)-f(y)\right],
\end{align}
and
\begin{eqnarray}
g(x) =
\left\{
 \begin{array}{ll}
\sqrt{\frac{1}{x}-1}\arcsin\sqrt{x} \;\;\;\;  & x\leq1 \\
\frac{\sqrt{1-\frac{1}{x}}}{2}\left(\log\frac{1+\sqrt{1-1/x}}{1-\sqrt{1-1/x}}-i\pi\right)  \;\;\;\; & x>1
 \end{array} 
\right. .
\end{eqnarray}

\section{Higgs trilinear couplings }\label{hhh}
 As discussed in Section~\ref{sec:4}, the partial decay width for $h_i \to h_j h_j$ ( $i>j$) is given by
\begin{equation}
 \Gamma(h_i\to h_j h_j)=\frac{g_{h_i h_j h_j}^{2}}{32\pi M_{h_{i}}}\left(1-\frac{4M_{ h_j}^{2}}{M_{h_{i}}^{2}}\right)^{1/2},
 \end{equation}
where the coupling $g_{h_i h_j h_j}$ can be expressed as
\begin{align}
 g_{h_2 h_1 h_1}&=\frac{1}{2} \left[ r_{13}^2 (\Lambda r_{21} v+r_{22} (-3
 \sqrt{2} \kappa _2+\sqrt{2} \kappa _3 \right. 
\nonumber\\ 
&+2 \lambda_s w_1)+ 6 \lambda_s r_{23} w_2)+r_{12}^2(\Lambda r_{21} v
\nonumber\\ 
& +3 r_{22} (\sqrt{2} (\kappa _2+\kappa_3)+2 \lambda_s w_1)+2 \lambda_s r_{23} w_2)
\nonumber\\ 
&+r_{11}^2 (3 \lambda r_{21}v + \Lambda ( r_{23}w_2 +r_{22} w_1))
\nonumber\\ 
&+2 \Lambda r_{11} (r_{13} (r_{23} v+ r_{21} w_2)
\nonumber\\ 
&+r_{12} (r_{22} v+r_{21} w_1
 ))+2 r_{12} r_{13} (r_{23} (-3 \sqrt{2} \kappa_2
\nonumber\\
& \left. +\sqrt{2} \kappa _3+2 \lambda_s w_1)+ 2
 \lambda_s r_{22} w_2) \right].
	 \end{align}
The $g_{h_3h_1h_1}$ coupling can be obtained from the above expression by substituting $r_{2j} \to r_{3j}$,
and for $g_{h_3h_2h_2}$ by the substitution of $r_{2j} \to r_{3j}$ and then $r_{1j} \to r_{2j}$.

\section{ Oblique parameters}\label{A-obli}
As discussed in Section \ref{sec:4}, the additional particles introduce corrections to the gauge boson propagators in the SM that can be parametrised by the oblique parameters $S$ and $T$~\cite{Peskin:1991sw}.
These parameters in the cSMCS may be written as
\begin{align}
T= \frac{g^2}{64 \pi^2 M_W^2 \alpha_{em}}&\bigg\{-(r_{12}r_{23}-r_{13} r_{22})^2 F(M_{h_1}^2,M_{h_2}^2)\nonumber
\\&-(r_{12} r_{33} -r_{13} r_{32} )^2F(M_{h_1}^2,M_{h_3}^2)\nonumber
\\&-(r_{22} r_{33} -r_{32} r_{32} )^2F(M_{h_2}^2,M_{h_3}^2)\nonumber
\\&+3r_{11}^2(F(M_Z^2,M_{h_1}^2)-F(M_W^2,M_{h_1}^2))\nonumber
\\&-3(F(M_Z^2,M_{h_{ref}}^2)-F(M_W^2,M_{h_{ref}}^2))\nonumber
\\&+3r_{21}^2(F(M_Z^2, M_{h_2}^2)-F(M_{W}^2,M_{h_2}^2))\nonumber
\\&+3r_{31}^2(F(M_Z^2, M_{h_3}^2)-F(M_{W}^2,M_{h_3}^2))
\bigg\},
 \label{T}
\end{align}
and
\begin{align}
S=\frac{g^2 }{384 \pi^2 \cos^2 \theta_W}&\bigg\{(r_{12}r_{23}-r_{13}r_{22})^2 G(M_{h_1}^2,M_{h_2}^2,M_Z^2)\nonumber
\\&+(r_{12} r_{13}-r_{13} r_{32} )^2 G(M_{h_1}^2,M_{h_3}^2,M_Z^2)\nonumber
\\&+(r_{22} r_{33}-r_{32} r_{32} )^2 G(M_{h_2}^2,M_{h_3}^2,M_Z^2)\nonumber
\\&+r_{11}^2\widehat{G}(M_{h_1}^2,M_Z^2)-\widehat{G}(M_{h_{ref}}^2,M_Z^2)\nonumber
\\&+r_{21}^2\widehat{G}(M_{h_2}^2,M_Z^2)+r_{31}^2\widehat{G}(M_{h_3}^2,M_Z^2)\nonumber
\\&+\log(M_{h_1})^2-\log(M_{h_{ref}})^2+\log(M_{h_2})^2\nonumber
\\&+\log(M_{h_3})^2 \bigg\},
 \label{S}
\end{align}
where the following functions have been used
\begin{equation}
F(M_1^2,M_2^2)=\frac{1}{2}(M_1^2+M_2^2)-\frac{M_1^2M_2^2}{M_1^2-M_2^2}\log(\frac{M_1^2}{M_2^2}),
\end{equation}
and
\begin{align}
G(m_1,m_2,m_3) &=\frac{-16}{3}+\frac{5(m_1+m_2)}{m_3}-\frac{2(m_1-m_2)^2}{m_3^2}\nonumber
\\& +\frac{3}{m_3} \left[\frac{m_1^2+m_2^2}{m_1-m_2}-\frac{m_1^2-m_2^2}{m_3} \right. \nonumber
\\&\left. +\frac{(m_1-m_2)^3}{3m_3^2}\right]\log\frac{m_1}{m_2}+\frac{r f(t, r)}{m_3^3}.
\label{G}
\end{align}
The function $f$ is given by
\begin{equation}
f(t, r)=\left\{\begin{array}{cc}
\sqrt{r} \ln|\frac{t-\sqrt{r}}{t+\sqrt{r}}| & r > 0\\
 0 & r = 0\\
2\sqrt{-r} \arctan\frac{\sqrt{-r}}{t} & r < 0
\end{array}
\right. ,
\end{equation}
with the arguments defined as
\begin{equation}
 t\equiv m_1+m_2-m_3 \, \,, \, \, r\equiv m_3^2-2m_3(m_1+m_2)+ (m_1-m_2)^2.
\end{equation}
Finally, $\widehat{G}(m_1,m_2)$ can be written as follows
\begin{align}
\widehat{G}(m_1,m_2)&=\frac{-79}{3}+9\frac{m_1}{m_2}-2\frac{m_1^2}{m_2^2} +\left( -10+18\frac{m_1}{m_2} \right.
\nonumber\\
& \left. -6\frac{m_1^2}{m_2^2}+\frac{m_1^3}{m_2^3}-9\frac{m_1+m_2}{m_1-m_2}\right)\log\frac{m_1}{m_2} 
\nonumber\\
&+(12-4\frac{m_1}{m_2}+\frac{m_1^2}{m_2^2}) \frac{f(m_1,m_1^2-4 m_1 m_2)}{m_2}. 
\label{'G'}
\end{align}

\bibliographystyle{JHEP}
\bibliography{bib_CHcol}{}

\bibliographystyle{unsrt}

\end{document}